\documentclass[12pt]{article}
\usepackage{amssymb}
\usepackage{amsmath}
\usepackage{graphicx}
\usepackage{rotating}
\usepackage{indentfirst}
\usepackage{cmap}
\usepackage{ulem}
\usepackage{color}
\usepackage{cite}
\usepackage{mathdots}
\usepackage{setspace}
\usepackage{mcite}

\begin{document}
\title{Spherically symmetric approaches\\ in the theoretical study\\ of low-dimensional magnets}
\author{A.\,F.\,Barabanov$^1$, V.\,E.\,Valiulin$^{1,2}$, A.\,V.\,Mikheyenkov$^{1,2}$\thanks{mikheen@bk.ru}, P.\,S.\,Savchenkov$^{3,4}$ \\
$^1$ Institute for High Pressure Physics, Russian Academy of Sciences, 108840, Moscow \\
$^2$ Moscow Institute of Physics and Technology, 141701, Dolgoprudny, Moscow region \\
$^3$ National Research Center Kurchatov Institute, 123182, Moscow \\
$^4$ National Research Nuclear University MEPhI, 115409, Moscow}

\maketitle

\begin{abstract}

The main ideas and some of the most important results of the spherically symmetric self-consistent approach and a number of related theoretical algorithms are presented. These methods make it possible to study low-dimensional Heisenberg-type spin models, including frustrated ones, with careful consideration of the theoretic (Mermin-Wagner and Marshall) theorems, as well as the site spin constraint. Thus, the difficulties that may arise in the traditional analysis of low-dimensional magnetic systems are avoided. The approach can also be applied to the spin-pseudospin model, and is also embedded in more complex constructions when considering spin models with free carriers, such as the basic and three-band Hubbard models, \(t-J\) and \(s-d\) models, and the Kondo lattice.
\end{abstract}

\renewcommand{\contentsname}{Contents}
\begin{spacing}{0.7}
\small{\tableofcontents}
\end{spacing}

\section{Introduction}

In low-dimensional systems, including magnetic ones, the role of quantum fluctuations is significant. Frustration serves as an additional disruptive factor. Frustration in compounds with localized magnetic moments can lead to interesting and often nonintuitive effects. For example, it can lead to the formation of a nontrivial spin configuration or the loss of spin ordering even at zero temperature. Like the role of quantum fluctuations, frustration effects are particularly strong in low-dimensional, particularly two-dimensional, systems.

The description of these effects within the framework of traditional methods of magnetism theory can be difficult, which prompts the search for alternative algorithms. One such approach is the spherically symmetric self-consistent approach (SSCA) (or, in Western terminology, rotation-invariant Green’s function method (RGM)).

Kondo and Yamaji proposed the RGM idea for the 1D Heisenberg model half a century ago in their pioneering work \cite{Kondo72_PTP}. The idea is conceptually simple: in the chain of equations for two-time retarded spin-spin Green's functions, decoupling is performed not at the first step (as is traditionally done \cite{Tyabli67_Book_Ru}), but at the second step. This scheme allows the spin state to be described not through the average spin, which can now be zero, but through spin-spin correlators\footnote{The idea of extracting spin-spin correlators from the Green's function was proposed in \cite{Callen63_PR} long before Kondo and Yamaji, but there it occurred at the first step after algebraic manipulations.}.

The next impetus for the development was given after the advent of high-temperature superconducting cuprates: in almost simultaneously published papers \cite{Shimah91_JPSJ} and \cite{Baraba92_JPSJ}, the Kondo-Yamaji method was extended to the two-dimensional square Heisenberg lattice, which is relevant for cuprates. Almost immediately, a version for a frustrated lattice was developed with exchanges on the first and second neighbors \cite{Baraba94_JPSJ,Baraba94_JETP_R}.

Subsequently, the RGM was repeatedly modified and supplemented, and was also extended to more complex models. Initially, of course, all these developments were driven by the search for a theory of high-temperature superconductivity (HTSC). Later, this self-limitation was lifted, and now this area of research can be considered quite independent and fully formed.

Below, the main ideas and some of the most important results of the RGM are presented, primarily using the example of a two-dimensional square lattice.

\section{Preliminary remarks}
\subsection{Space dimensionality}

\subsubsection{3D and 1D}
The RGM provides a reasonable description of the paramagnetic state of a magnetic system (the average spin at a site is zero, \(\langle \mathbf{\hat{S}}_{\mathbf{i}}\rangle = 0\)) with strong spin-spin correlations. Therefore, it is applicable, for example, in the three-dimensional case at temperatures slightly above the Curie or Neel temperature, of course, outside the fluctuation region\footnote{In fact, the RGM works in a wide temperature range in 3D, more on this in Section \ref{to_3D}.}.

In the opposite (one-dimensional) limit, the use of the RGM is meaningless due to the existence of an exact solution. However, the RGM can be used for frustrated or so-called decorated one-dimensional systems and for several weakly coupled chains, where there is no exact solution \cite{Hartel08_PRB,Richte09_JPCS,Hartel11_PRB,Hutak22_EPJB,Hutak23_EPJB}. The same applies to several weakly coupled planes \cite{Schmal05_PRB,Schmal06_PRL,Kozlov07_JL_R}.

\subsubsection{2D}
A natural domain of application for the RGM is purely two-dimensional systems. Here, two lines of research have traditionally existed: the `line of exact theorems' and the `line of weak three-dimensionality'.

In the first case, the basic thesis is as follows: the Mermin-Wagner theorem \cite{Mermin66_PRL} in 2D forbids spontaneous symmetry breaking at \(T > 0\); i.e., the familiar picture of a magnet with an average spin \(\langle \mathbf{\hat{S}}_{\mathbf{i}}\rangle \neq 0\), directed either in one or in alternating directions, is inadequate. Furthermore, the Marshall theorem \cite{Marsha97_PRSLSA} requires a singlet ground state for the antiferromagnetic phase. Accordingly, standard analytical methods applicable in 3D must be modified to account for these constraints.

The second approach is based on the thesis: "strictly two-dimensional systems do not exist." Weak three-dimensionality effects are always present, which automatically rules out 2D restrictions. In the worst case, we are dealing not with long-range order, but with an exponentially large correlation length. The Marshall theorem, which is only valid at zero temperature, can be neglected entirely. Therefore, standard methods need only be modified to the extent required by the specific geometry (number of nearest neighbors, filling of coordination spheres, etc.).

The debate between these two concepts, which was quite heated during the early HTSC boom, subsequently died down, and now they peacefully coexist. Some of the basic properties of a 2D magnetic system, calculated within the framework of either concept, coincide qualitatively and semiquantitatively. However, there are also significant differences. An analysis of these differences would take too much space and is not the purpose of this paper.

The RGM belongs to the first approach. In this approach, the state of the magnetic system for any number and any sign of exchange interactions, at any temperature \(T \geq 0\), is singlet, all sites are equivalent, the average spin at a site is zero \(\langle \mathbf{\hat{S}}_{\mathbf{i}}\rangle = 0\), and the spin (long-range or short-range) order is described by spin-spin correlators.

Thus, in the RGM, neither spin nor lattice symmetry is broken.

\subsection{Lattice geometry}
Thus, a natural application area of the RGM is the 2D spin lattice. The next issue concerns the specific geometry of this lattice.

Lattices with geometric frustration occupy a special place, i.e., those in which the antiferromagnetic checkerboard pattern is impossible (triangular lattices, Kagome lattices, and in 3D, these are pyrochlore and hyper-Kagome lattices, etc.). In systems of this kind, the concept of chirality (the difference between right and left) and a host of other striking effects, including skyrmions, naturally arise. Research in this area is extensive and of considerable interest (see, for example, reviews \cite{Staryk15_RPP,Zhou17_RMP,Markin21_UFN_R} and the references therein). Here, the RGM has proven itself to be a quite effective method (triangular lattices \cite{Rubin05_PLA,Rubin08_PLA,Antsyg08_PRB,Rubin10_PLA,Rubin12_PLAa,Rubin17_JPCS}, Kagome lattices, including quasi-three-dimensional ones \cite{Yu00_EPJB,Bernha02_PRB,Schmal05_PRB, Mueller18_PRB}, hexagonal lattices, including quasi-three-dimensional ones \cite{Vladim17_EPJB, Vladim18_EPJB}, and pyrochlore lattices \cite{Mueller17_PRB, Mueller19_PRB}).

However, we will focus on another case — a square lattice (note that frustration also arises in this lattice, but for different reasons). Numerous compounds with weakly coupled square (or reducible to square planes via superexchange) magnetic planes have been experimentally studied. The best-known and best-studied are the HTSC cuprates, but, we repeat, they are by no means the only ones.

As for lattices with geometric frustration, this area deserves separate consideration, and even a brief analysis would increase the scope of this review to an unacceptable level.

\subsection{Spin magnitude}
Thus, below we consider the application of the RGM using the example of a 2D square spin lattice. A final preliminary remark concerns the spin magnitude.

In principle, the method works for any spin (see, for example, \cite{Junger08_PRB,Suzuki94_JPSJ,Mueller17_PRB,Mueller18_PRB}). However, as can be shown, for the most interesting antiferromagnetic case, calculations within the RGM framework for spin \(S > 1\) lead to the following: the gap in the spin excitation spectrum at the antiferromagnetic point of the Brillouin zone \(\mathbf{Q}=(\pi ,\pi)\) is exponentially small up to temperatures on the order of the exchange interaction. Accordingly, the correlation length is exponentially large. Although the state of the system in the RGM in this case remains singlet, all calculated quantities — the spectrum of spin excitations, the structure factor, and the magnetic susceptibility — are almost identical to those obtained using the standard spin-wave approach or approaches close to it. Thus, for large spins, the RGM, although applicable, appears to be redundant.

The case of \(S = 1\) is intermediate and is also studied within the RGM framework for various lattices \cite{Suzuki94_JPSJ,Junger05_PRB,Rubin17_JPCS,Rubin10_PLA,Rubin12_PLAa,Mueller17_PRB,Mueller18_PRB} (in some lattices, the case of arbitrary spin is also analyzed \cite{Mueller18_PRB,Mueller17_PRB}). However, the vast majority of studies consider \(S = 1/2\), where the differences between the RGM results and those of the spin-wave approach (or approaches close to it) are often significant.

\section{\(J_1\)-\(J_2\)-\(J_3\) model, a sequence of complications}
\subsection{Hamiltonian}
Thus, we consider a two-dimensional square lattice, with each site assigned a spin of \(S=1/2\) (exceptions will be discussed in Sections \ref{Kug_Khom}, \ref{Spin_polar} and \ref{to_3D}).

\begin{figure}[h]
  \centering
  \textbf{a} \includegraphics[width=0.4\textwidth]{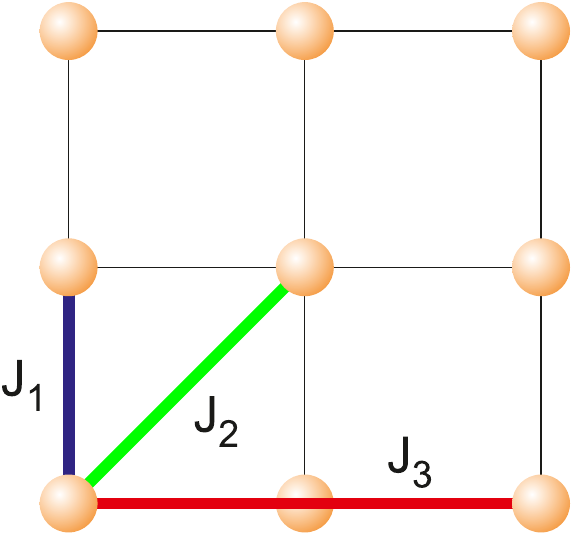} \hspace{1cm}
  \textbf{b} \includegraphics[width=0.4\textwidth]{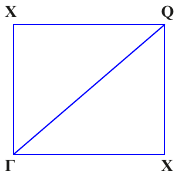}
  \caption{\small \textbf{a.} (Color online) First, second, and third nearest neighbors for a 2D square lattice and the corresponding exchange parameters.
    \textbf{b.} Standard notations for symmetric points in the Brillouin zone of a square lattice.
  \(\mathbf{\Gamma} = (0,0)\) is the zero (FM) point, \(\mathbf{X} = (0,\pi), (\pi,0)\) are stripe points,
  \(\mathbf{Q} = (\pi,\pi)\) is the AFM point. A quarter of the full zone is shown.}
  \label{fig_Neighbours}
\end{figure}

The Hamiltonian has the standard Heisenberg form, and three cases are examined: interactions between nearest neighbors only, nearest neighbors and next-nearest, neighbors on the third coordination sphere (see Fig.~\ref{fig_Neighbours}a). Thus, the Hamiltonian has the form:
\begin{equation}
\hat{H} = J_{1}\sum_{\langle\mathbf{i},\mathbf{j}\rangle}
\hat{\mathbf{S}}_{\mathbf{i}}\hat{\mathbf{S}}_{\mathbf{j}}+
J_{2}\sum_{[\mathbf{i},\mathbf{j}]}\hat{\mathbf{S}}_{\mathbf{i}}
\hat{\mathbf{S}}_{\mathbf{j}}+
J_{3}\sum_{\{\mathbf{i},\mathbf{j}\}}\hat{\mathbf{S}}_{\mathbf{i}}
\hat{\mathbf{S}}_{\mathbf{j}}
\label{Hamilt}
\end{equation}
Here \(\left\langle \mathbf{i,j}\right\rangle\) denotes summation over pairs (bonds) of nearest-neighbor sites, \(\left[ \mathbf{i,j}\right]\) over second-neighbor pairs, \(\left\{ \mathbf{i,j}\right\}\) over third-neighbor pairs; as already mentioned, \((\mathbf{\hat{S}}_{\mathbf{i}})^{2}=3/4\).

\subsection{Sequence of complications}
\label{Seq_compl}
The choice of a model with three exchanges is not random. At each subsequent step of increasing the complexity of the model, qualitatively new effects arise. Looking ahead, we will describe them briefly (details and references will be presented in Sections \ref{J1}, \ref{J1-J2} and \ref{J1-J2-J3}).

\textit{Model \(J_1\).}
A two-dimensional antiferromagnet (AFM) and ferromagnet (FM) are described in exact accordance with the Marshall \cite{Marsha97_PRSLSA} and Mermin-Wagner \cite{Mermin66_PRL} theorems. The state of the system at \(T \geq 0\) is a singlet with zero average spin per site. There is no spontaneous symmetry breaking in spin space. In direct space, translational invariance and lattice symmetry are preserved.

At \(T = 0\) long-range order arises: the modulus of the spin-spin correlator at infinity \(\langle \mathbf{\hat{S}}_{\mathbf{0}}\mathbf{\hat{S}}_{\mathbf{R}}\rangle (R\rightarrow \infty )\) reaches a nonzero asymptotic value, the sign of the correlator is constant for the FM case and alternates in a checkerboard pattern in the AFM case.

The spin constraint condition \(\langle \mathbf{\hat{S}}_{\mathbf{i}}^{2}\rangle = 3/4\) is satisfied at each site exactly\footnote{This dual-language term (constraint) has actually become commonly used in the RGM.}. Note that in the standard spin-wave approach and similar approaches, the need to satisfy this condition is usually simply ignored; in some other approaches, for example, in the auxiliary boson method \cite{Auerba11_Book_Chap}, the constraint is satisfied only on average over the entire system.

\textit{Model \(J_1\)-\(J_2\).}
All of the above mechanisms are valid. At \(T = 0\), three phases with long-range order are realized: AFM, FM, and stripe phase, as well as two disordered spin-liquid phases. The loss of long-range order upon transition to a spin-liquid state is a canonical example of a quantum phase transition \cite{Sachde11_Book}. At \(T > 0\) and for any \(J_1\) and \(J_2\), long-range order is absent; short-range order evolves in accordance with the structure of the zero-temperature phase.

\textit{Model \(J_1\)-\(J_2\)-\(J_3\).}
In addition to the phases listed above, three different helical spin phases appear. It is significant that the helices arise without the Dzyaloshinsky-Moriya interaction \cite{Dzyalo57_JETP_R,Dzyalo58_JPCS,Moriya60_PRL,Moriya60_PR} (see also the recent review \cite{Boriso20_UFN_R}).

Furthermore, within a certain range of parameters, a state with two interpenetrating long-range orders is realized.

Finally, a model with more than three exchanges is difficult to justify physically. For the three previous cases considered above, physical realizations do exist.

Let us emphasize once again that the different phases and, accordingly, the phase transitions between them in the RGM occur only at \(T = 0\). At \(T > 0\), the state of the system is always paramagnetic — albeit with strong correlations, and their structure depends on the exchange parameters.

As a side remark, note that in a triangular lattice, despite its completely different geometry, the effects arising with an increase in the number of exchanges are qualitatively similar. At \(T = 0\), the inclusion of \(J_2\) leads to new long-range-order structures as well as disordered regions; at \(T > 0\), the number of distinct short-range-order structures increases. The third exchange \(J_3\) generates incommensurate helicoidal structures \cite{Rubin10_PLA,Rubin12_PLAa,Rubin17_JPCS}.

\section{Magnetic order in the classical limit}
\subsection{Emergence of new phases with increasing number of exchanges}
\label{Class_phases}
Before moving to the ultraquantum limit \(S=1/2\), it is natural to consider the opposite (classical) limit \(S \gg 1\).

In this limit, on a 2D square lattice, any long-range order — commensurate or incommensurate — is described by the standard ansatz (plane spiral) \cite{Luttin46_PR,Diep13_WS}
\begin{equation}
\mathbf{S_r} = \mathbf{e_1}\cos(\mathbf{q}_0 \mathbf{r}) + \mathbf{e_2}\sin(\mathbf{q}_0 \mathbf{r}),
\label{S_r}
\end{equation}
where \(\mathbf{e_1}\) and \(\mathbf{e_2}\) are orthogonal unit vectors lying in the plane (the spin length is normalized to unity), \(\mathbf{r}\) is the site position vector. The point \(\mathbf{q}_0\) in the Brillouin zone — it is this point that determines the magnetic order — is sometimes called the control point. A quarter of the Brillouin zone of the square lattice with the accepted notation of symmetric points is shown in Fig.~\ref{fig_Neighbours}b.

For given values of the exchange parameters \(J_1, J_2, J_3\), the spin structure is found by substituting (\ref{S_r}) into (\ref{Hamilt}), where instead of spin operators there are classical unit vectors, and by minimizing the classical energy with respect to the position of the control point \(\mathbf{q}_0\).

\textit{For the \(J_1\) model} with normalized exchange parameter \(\left\vert J_{1}\right\vert = 1\), the set of exchange parameter values consists of two points \(J_{1} = \pm 1\). Substitution (\ref{S_r}) gives a trivial result: for \(J_1 > 0\) the AFM phase is realized (control point \(\mathbf{q_0} = \mathbf{Q}=(\pi ,\pi)\)\footnote{Sometimes in the literature, the notation \(\mathbf{M}\) is used for this point.}), for \(J_1 < 0\) the FM phase is realized (control point \(\mathbf{q_0} = \mathbf{\Gamma }=(0,0)\)).

\begin{figure}[h]
  \centering
  \textbf{a} \includegraphics[width=0.41\textwidth]{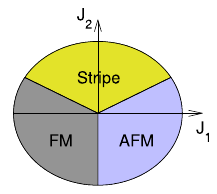}
  \textbf{b} \includegraphics[width=0.4\textwidth]{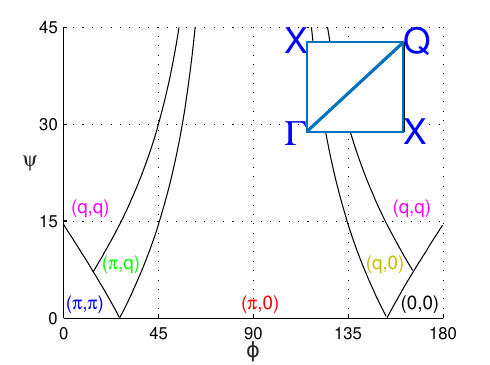}
  \caption{\small (Color online) \textbf{a.} Phase diagram of the \(J_1\)-\(J_2\) circle for a square lattice in the classical limit. The AFM, stripe, and FM phases are realized. Transitions to the stripe phase occur at \(0 < J_2 = \pm J_1/2\).\\
  \textbf{b.} Flat scan of part of the classical \(J_1\)-\(J_2\)-\(J_3\) sphere phase diagram, \(0 \leq \varphi \leq \pi\), \(0 \leq  \psi \leq  \pi/2\). The positions of the control points are shown. Inset: a quarter of the Brillouin zone.}
  \label{fig_Circle_Sph_CL}
\end{figure}

\textit{For the \(J_1\)-\(J_2\) model} it is natural to use the parameterization
\begin{equation}
J_{1} = \cos \varphi, \quad J_{2} = \sin \varphi,\quad  0 \leq \varphi \leq 2\pi,
\quad \sqrt{J_{1}^{2}+J_{2}^{2}} = 1.
\label{Param_12}
\end{equation}
Then the set of exchange parameter values is a circle of unit radius. Substitution (\ref{S_r}) leads to slightly more complicated equations, the solution of which shows that, along with the AFM and FM phases, a stripe phase appears. It represents an alternation of stripes with spin up and spin down. This state is doubly degenerate in accordance with the two possible directions of the stripes. The control points for the stripe phase are \(\mathbf{q_0} = \mathbf{X}=(0,\pi ),(\pi ,0)\). The regions of existence of the three phases are shown in Fig.~\ref{fig_Circle_Sph_CL}a.

\textit{Now the \(J_1\)-\(J_2\)-\(J_3\) model.}
Here a `spherical' parameterization is convenient\footnote{For studying local parts of the parameter region, parameterizations different from (\ref{Param_12}) and (\ref{Param_123}) are often convenient. However, for the full regions of exchanges, (\ref{Param_12}) and (\ref{Param_123}) seem most adequate.}
\begin{eqnarray}
J_{1} &=&\cos (\psi )\cos (\varphi ),\quad J_{2}=\cos (\psi )\sin (\varphi
),\quad J_{3}=\sin (\psi ),\   \label{Param_123} \notag \\
0 &\leq &\varphi \leq 2\pi ,\quad -\pi/2\leq \psi \leq \pi/2,\quad \sqrt{%
J_{1}^{2}+J_{2}^{2}+J_{3}^{2}}=1
\end{eqnarray}
The set of exchange parameter values is a sphere of unit radius.

Substitution (\ref{S_r}) generates rather cumbersome equations, the solution of which leads to the appearance of three more phases. Below are all possible positions of the control point and the corresponding values of the classical energy.
\begin{equation}
\begin{array}{ccccccc}
\vspace{3pt}
& \mathbf{q}_{0} & E(\mathbf{q}_{0}) & \qquad  &  & \mathbf{%
q}_{0} & E(\mathbf{q}_{0}) \\
\vspace{3pt}
\mathbf{1} & (0,0) & 4(J_{1}+J_{2}+J_{3}) & \quad  & \mathbf{4} &
(0,q_{1}),(q_{1},0) & 2J_{1}-\frac{(J_{1}+2J_{2})^{2}}{4J_{3}} \\
\vspace{3pt}
\mathbf{2} & (\pi ,\pi ) & -4(J_{1}-J_{2}-J_{3}) & \quad  & \mathbf{5} &
(\pi ,q_{2}),(q_{2},\pi ) & -2J_{1}-\frac{(J_{1}-2J_{2})^{2}}{4J_{3}} \\
\vspace{3pt}
\mathbf{3} & (0,\pi ),(\pi ,0) & -4(J_{2}-J_{3}) & \quad  & \mathbf{6} &
(q_{3},q_{3}) & -4J_{3}-\frac{J_{1}^{2}}{J_{2}+2J_{3}}\\
\label{Class_Ener_123}
\end{array}
\end{equation}
where the notation is:
\begin{equation}
q_1 = \arccos(-\frac{J_1+2J_2}{4J_3}), \quad
q_2 = \arccos(-\frac{J_1-2J_2}{4J_3}), \quad
q_3 = \arccos(-\frac{J_1}{2J_2+4J_3})
\label{Class_helix}
\end{equation}

In expressions (\ref{Class_Ener_123}), the first three phases (in the left column) are the already mentioned FM, AFM, and stripe structures. Phases 4–6 (in the right column) are spin helices (helicoidal phases); the spins rotate from site to site, and the period of the spin structure is not necessarily commensurate with the lattice period.

Figure~\ref{fig_Circle_Sph_CL}b shows a flat scan of the most interesting quarter of the full sphere — the phase diagram \(0 \leq \varphi \leq \pi\), \(0 \leq  \psi \leq  \pi/2\), i.e., \(1 \geq J_1 \geq -1\), \(0 \leq J_2 \leq 1\), \(0 \leq  J_3 \leq  1\) (angles are shown in degrees\footnote{Parametric angles are shown in degrees to avoid confusion with coordinates — in radians — in the Brillouin zone.}). All experimental data on quasi-two-dimensional compounds known to the authors fall within this region.

\subsection{Features of the classical phase diagram}
The following features of the phase diagram are noteworthy.

\textbf{i.} For FM, AFM, and stripe phases, the control point is stationary throughout the entire region of existence of the corresponding phase and abruptly changes position upon transition to another phase. For helicoids, however, as the exchange parameters change, point \(\mathbf{q}_0\) moves along the symmetric lines of the Brillouin zone in accordance with (\ref{Class_helix}). For helicoids Nos~4–5 from expressions (\ref{Class_Ener_123}), this movement is along the sides of the Brillouin zone, while for helicoid No.~6, it is diagonal. Thus, helicoids with a varying pitch perform a continuous transition between symmetric points.

We emphasize that, as mentioned above, in the \(J_1\)-\(J_2\)-\(J_3\) model, spin helices arise already in the classical limit, without breaking inversion symmetry and without invoking the Dzyaloshinsky-Moriya interaction.

\textbf{ii.} The entire region of exchange parameters can be divided into frustrated and nonfrustrated ones. In the nonfrustrated region, all exchanges operate in one direction, and all bonds are fully saturated. The simplest case is \(J_1,J_2,J_3 < 0\), when all three exchanges favor the FM phase. Another example is \(J_1 > 0\), \(J_2 < 0\), \(J_3 > 0\), with all bonds in the AFM configuration being saturated.

The literature discusses the types and extent of frustration \cite{Diep13_WS} and (in the quantum case) its connection with entanglement \cite{Wolf03_IJQI,Amico08_RMP,Facchi10_NJP}, but we will not pursue this direction further here.

The quadrant of the full phase diagram shown in Fig.~\ref{fig_Circle_Sph_CL}b is maximally frustrated — in the sense that if at least two exchanges are nonzero, then all bonds in Hamiltonian (\ref{Hamilt}) cannot be simultaneously saturated. In the remaining three quadrants, this is not true for at least one of the bonds. Frustration leads to the appearance of an intermediate stripe phase between the AFM and FM phases, and, for \(J_3>0\), also to the appearance of helicoids. In the quantum case, the consequences of frustration are even more profound (see below).

\textbf{iii.} The question of the experimental feasibility or fundamental nonfeasibility of specific states can be discussed. It is clear, for example, that the case \(J_3 \gg J_1,J_2\) is unlikely to ever be observed experimentally. In this sense, it is significant that some helicoids arise even at small \(J_3\) values. In general, however, this question does not admit a definitive answer.

\subsection{On the quantum phase diagram}

Thus, in the classical limit of the \(J_1\)-\(J_2\)-\(J_3\) model, six spin configurations are possible: FM, AFM, stripe, and three different helices.

Within the framework of the RGM, in the ultraquantum limit \(S=1/2\) at \(T=0\), the situation is qualitatively similar — all these long-range-order phases are possible (although the long-range order itself is described not by the average site spin, but by spin-spin correlators). However, even at \(T=0\) states without long-range order (spin liquids) may also occur. In the phase diagram, they arise between long-range-order phases in the presence of frustration.

At \(T \neq 0\), long-range order is impossible; short-range order evolves with changing exchange parameters in accordance with the structure of zero-temperature phases. A detailed description is given below in Section \ref{Jall}.

\section{Algorithm of the spherically symmetric self-consistent approach}
\label{Algor}
\subsection{Calculation scheme}
\label{calc_scheme}

Let us return to Hamiltonian (\ref{Hamilt}). The primary calculated quantity in the RGM is the double-time retarded site spin-spin Green's function \cite{Zubare60_UFN_R}
\begin{equation}
G_{\mathbf{nm}}^{zz} = \langle S_{\mathbf{n}}^{z}|S_{\mathbf{m}}^{z}\rangle
_{\omega +i\delta } = -i\int\limits_{0}^{\infty }dt\,e^{i\omega t}\langle
\lbrack S_{\mathbf{n}}^{z}(t),S_{\mathbf{m}}^{z}]\rangle
\label{GreenF}
\end{equation}

Since a spherically symmetric state is considered, only the diagonal Green's functions \(G^{zz}=G^{xx}=G^{yy}\) in the components \(\alpha =x,y,z\) are nonzero (and equal to each other); there are three branches of spin excitations degenerate in \(\alpha\), and the average spin at the site is zero \(\langle S_{\mathbf{n}}^{\alpha }\rangle =0\).

The main idea of the RGM algorithm is to close the chain of equations of motion for the Green's function (\ref{GreenF}) at the second step. The first equation of motion in the frequency representation has the form
\begin{equation}
\omega G_{\mathbf{nm}}^{zz} = \langle \left[ S_{\mathbf{n}}^{z},S_{\mathbf{m}}^{z}%
\right] \rangle +\langle [ S_{\mathbf{n}}^{z},\hat{H}] \mid
S_{\mathbf{m}}^{z}\rangle _{\omega +i\delta }
\label{First step}
\end{equation}

Since \(\langle \left[ S_{\mathbf{n}}^{z},S_{\mathbf{m}}^{z}\right] \rangle =0\), closure according to the traditional Tyablikov scheme at the first step \cite{Tyabli67_Book_Ru}, with the extraction of the average spin from the two-site Green's function, is impossible\footnote{Instead of \(\langle S_{\mathbf{n}}^{z}|S_{\mathbf{m}}^{z}\rangle\), one can consider the Green's function \(\langle S_{\mathbf{n}}^{+}|S_{\mathbf{m}}^{-}\rangle\) common for spin problems; then the requirement that the first term on the right-hand side of (\ref{First step}) vanish, \(\langle \lbrack S^{+},S^{-}]\rangle = 2\langle S^{z}\rangle\), is imposed directly.}.

The nonzero term on the right-hand side of (\ref{First step}) is the sum of several two-site Green's functions. Without revealing the specific form for now, we write the equation of motion for it.
\begin{equation}
\omega \langle \lbrack S_{\mathbf{n}}^{z},\hat{H}]|S_{\mathbf{m}}^{z}\rangle_{\omega +i\delta } =
\langle \lbrack \lbrack S_{\mathbf{n}}^{z},\hat{H}],S_{\mathbf{m}}^{z}]\rangle +
\langle \lbrack \lbrack S_{\mathbf{n}}^{z},\hat{H}],\hat{H}]|
S_{\mathbf{m}}^{z}\rangle_{\omega +i\delta}
\label{Second step}
\end{equation}

Here, the first term on the right-hand side is already nonzero, and the second term is the sum of various three-site Green's functions. From these, spin-spin correlators are extracted with closure to the original Green's function (\ref{GreenF}) (see (\ref{Vertex}) below for more details). The calculations take into account the relations arising from spherical symmetry: \(\langle S_{\mathbf{n}}^{x}S_{\mathbf{n+r}}^{x}\rangle = \langle S_{\mathbf{n}}^{y}S_{\mathbf{n+r}}^{y}\rangle = \langle S_{\mathbf{n}}^{z}S_{\mathbf{n+r}}^{z}\rangle = c_{\mathbf{r}}= c_{|\mathbf{r|}}\), where \(c_r\) is the spin-spin correlator\footnote{To avoid misunderstandings, we clarify that spherical symmetry is in spin space; in direct space, the geometry can be very different.} at distance \(r\).

After the Fourier transform
\begin{equation}
S_{\mathbf{q}}^{z} = \frac{1}{\sqrt{N}}\sum_{\mathbf{r}}e^{-i\mathbf{qr}}
S_{\mathbf{r}}^{z}
\end{equation}
the solution to the system of equations (\ref{First step})–(\ref{Second step}) gives the final expression for function (\ref{GreenF}).
\begin{equation}
G^{z}(\mathbf{q},\omega ) =
\langle S_{\mathbf{q}}^{z}|S_{-\mathbf{q}}^{z}\rangle_{\omega} =
-\chi(\mathbf{q},\omega) =
\frac{F_{\mathbf{q}}}{\omega ^{2}-
\omega _{\mathbf{q}}^{2}}.  \label{Gz}
\end{equation}
(hereafter, the second superscript \(z\) on the Green's function is omitted, \(\chi(\mathbf{q},\omega)\) is the susceptibility).

The final expressions for the numerator \(F_{\mathbf{q}}\) and the spectrum of spin excitations \(\omega _{\mathbf{q}}^{2}\) depend on the lattice geometry and the form of the Hamiltonian. But in any case, they contain spin-spin correlators on the first few coordination spheres. From simple geometric considerations, it is clear that for a square lattice in the \(J_1\) model, there are three such coordination spheres, for \(J_1\)-\(J_2\) there are five, and for the \(J_1\)-\(J_2\)-\(J_3\) model eight coordination spheres are involved. Cumbersome expressions for \(F_{\mathbf{q}}\) and \(\omega _{\mathbf{q}}^{2}\) in the most general case are given in the Appendix.

These correlators are expressed in the standard way through \(G^{z}(\mathbf{q},\omega )\)
\begin{equation}
c_{\mathbf{r}} = \langle S_{\mathbf{n}}^{z}S_{\mathbf{n+r}}^{z}\rangle =
\frac{1}{N}\sum_{\mathbf{q}}c_{\mathbf{q}}e^{i\mathbf{qr}}
\label{cq}
\end{equation}
\begin{equation}
c_{\mathbf{q}} =
\left\langle S_{\mathbf{q}}^{z}S_{\mathbf{-q}}^{z}\right\rangle =
-\frac{1}{\pi }\int_{0}^{\infty }\!d\omega \left(2m(\omega )+1\right)
\mathrm{Im}\ G^{z}\left( \mathbf{q},\omega \right)
\label{cg_Gz}
\end{equation}
where \(m(\omega)\) is the Bose function at a given temperature. Recall that the Bose function automatically arises in the equations for the commutator Green's function, which in itself does not imply the presence of Bose-like excitations.

Also included is the intrasite correlator \(c_{\mathbf{r}= 0}\) — this is the spin constraint condition \(c_{\mathbf{r}=0} = \langle S_{\mathbf{n}}^{z}S_{\mathbf{n}}^{z}\rangle = 1/4\).

The resulting system of self-consistent equations is then solved numerically. Note that the required computational resources are incomparably smaller than those needed for the initial first-principles numerical solution, even for a small lattice fragment.

\textit{A necessary clarification to avoid misunderstandings.}
In the RGM, at any temperature, any number, and any sign of exchanges (including for FM), the state of the system is spherically symmetric, a singlet one. Strictly speaking, there is no mathematically impeccable proof of singletism in the RGM, since the state is not explicitly constructed. However, for any other state, it is difficult to expect the conditions for equality of correlators along the three axes, given after formula (\ref{First step}), to be satisfied.

\subsection{Case \(T = 0\)}

In the RGM, the zero-temperature case requires separate consideration. We will demonstrate this using the example of the AFM phase (the details vary for different phases).

At \(T = 0\), long-range order arises in the system. Mathematically, this is because at \(T \to 0\), a \(\delta\)-shaped peak appears in the structure factor \(c_\mathbf{q}\) (\ref{cg_Gz}) at the corresponding symmetric point of the Brillouin zone, in this case at the AFM point \(\mathbf{Q}\). On the right-hand side of expression (\ref{cg_Gz}), the contribution without the Bose function is insufficient to satisfy the constraint condition. The correlator \(c_{\mathbf{r}}\) necessarily takes the form
\begin{equation}
c_{\mathbf{r}} = \frac{1}{N}\sum_{\mathbf{q}}c_{\mathbf{q}}e^{i\mathbf{qr}}+
c^{\mathrm{afm}}_{\mathrm{cond}}e^{i\mathbf{Qr}}
\label{cq_cond}
\end{equation}

Here, \(c^{\mathrm{afm}}_{\mathrm{cond}}\) is the modulus of the spin-spin correlator at infinity (the term spin condensate is used), the correlator itself changes sign in a checkerboard pattern
\begin{equation}
\langle S_{\mathbf{0}}^{\alpha }S_{\mathbf{r}}^{\alpha }\rangle^{\mathrm{afm}}_{\mathbf{r\rightarrow \infty }} = c^{\mathrm{afm}}_{\mathrm{cond}}(-1)^{n_{x}+n_{y}},\quad
\alpha =x,y,z, \quad
\mathbf{r} = n_{x}\mathbf{g}_{x}+n_{y}\mathbf{g}_{y}
\label{cr_cond}
\end{equation}

Otherwise, the calculation scheme remains the same; \(c^{\mathrm{afm}}_{\mathrm{cond}}\) becomes an additional self-consistency parameter.

The above reasoning may seem insufficiently rigorous. A more accurate justification can be found, for example, in \cite{Baraba11_TMP_R}.

Generally speaking, a detailed consideration of the \(T = 0\) case is necessary primarily for completeness, to demonstrate that the method also works at zero temperature, where `Bose condensation of spin excitations' occurs, or to solve specific problems, such as those concerning quantum phase transitions. Most essential conclusions, including phase boundaries with long-range order, can be obtained by extrapolating low-temperature results. Therefore, in what follows, the zero-temperature case will be occasionally omitted.

\subsection{Vertex corrections}
\label{Vertex_sec}

In the RGM, beginning with the pioneering papers \cite{Kondo72_PTP,Shimah91_JPSJ,Baraba92_JPSJ,Baraba94_JPSJ,Baraba94_JETP_R}, so-called vertex corrections have been introduced (the term is not directly related to diagrammatic technique). The introduction of the correction leads to renormalization of the spin-spin correlator when extracting it from the three-site Green's function.
\begin{equation}
\langle S_{\mathbf{l}}^{\alpha }S_{\mathbf{l+r}}^{\alpha }S_{\mathbf{n}}^{z}|
\ldots \rangle \rightarrow \alpha _{r}\langle S_{\mathbf{l}}^{\alpha}
S_{\mathbf{l+r}}^{\alpha }\rangle \langle S_{\mathbf{n}}^{z}|\ldots \rangle
= \alpha_{r} c_{r}\langle S_{\mathbf{n}}^{z}|\ldots \rangle =
\tilde{c}_{r}\langle S_{\mathbf{n}}^{z}|\ldots \rangle
\label{Vertex}
\end{equation}
In all expressions where such extraction occurs, the renormalized correlators \(\tilde{c}_{r}\) are used (details can be found in the cited literature).

Vertex corrections depend on the lattice vector \(r\), and in principle, each coordination sphere can have its own \(\alpha_{r}\) (three, five, or eight for the \(J_1\), \(J_1\)-\(J_2\), and \(J_1\)-\(J_2\)-\(J_3\) models, respectively). The simplest single-vertex configuration (all vertices coincide) is sufficient to satisfy the constraint condition, and even in this case, all the main results are reproduced.

More complex adjustments are made based on various reasonable physical considerations; details can be found in the aforementioned papers. With increasing temperature, all corrections \(\alpha \to 1\). At low temperatures, in most studies, the vertex corrections do not deviate far from unity, \(\alpha \lesssim 1.5\).

The presence of vertex corrections should be kept in mind when comparing results from different studies. For the same model, calculated quantities always agree qualitatively and semiquantitatively, but may differ strictly quantitatively due to different vertex selection.

\section{Advantages and disadvantages of the spherically symmetric self-consistent approach}
\label{pro_contra}

The previous sections described the spherically symmetric self-consistent approach for double-time retarded spin Green's functions, whose main application is low-dimensional magnetic systems. We now list the main advantages and disadvantages of the RGM. Let us start with the disadvantages.

\textit{Disadvantages.}

\textbf{i.} Despite some sophistication and cumbersomeness, the RGM is still a mean-field approach. It does not allow one to determine the damping of spin excitations within its basic formulation. Progress in this direction is only possible through semi-phenomenological renormalizations or other complications.

\textbf{ii.} The RGM is a method with uncontrollable accuracy. Although the error can often be estimated using various secondary considerations, such estimates are not intrinsic to the method. As a rule, there is no small parameter (however, this situation is common for magnetic problems).

\textbf{iii.} The method includes one or more tuning parameters (vertex correction(s)).

\textbf{iv.} The studied state is not structurally constructed; it is described via the Green's function, spin correlators, structure factor, and excitation spectrum. Although, strictly speaking, this can hardly be considered a significant drawback.

\textbf{v.} A local limitation arises in the spin-liquid region (generally, there are several, but the one located between the AFM and stripe phases has been better studied, see Fig.~\ref{fig_Circle_Quant}). Methods based on resonating valence bond (RVB) concepts, depending on the exchange parameters, reveal multiple structures here, differing in higher-order correlators (see, for example, \cite{Savary16_RPP,Zhou17_RMP} and references therein), whereas the RGM describes this region as a single phase without long-range order with evolving short-range order.

\textit{Advantages.}

\textbf{i.} The RGM satisfies the Mermin-Wagner and Marshall theorems, as well as the exact spin constraint at each site.

\textbf{ii.} At \(T = 0\), the method allows one to analyze states with and without long-range order.

\textbf{iii.} Over a wide temperature range \(T > 0\), a single approach reproduces all possible short-range-order structures in the system.

\textbf{iv.} The RGM allows one to calculate: the spin excitation spectrum \(\omega_{\mathbf{q}}\), the dynamic susceptibility \(\chi(\mathbf{q},\omega,T)\), the structure factor \(c_{\mathbf{q}}\), energy, and heat capacity. Crucially, no initial assumptions are made about the form of the spin excitation spectrum; the spectrum is determined through self-consistency.

\textbf{v.} The method can be embedded into more complex constructions when considering spin models with free carriers, such as the basic and three-band Hubbard models, the \(t-J\) and \(s-d\) models, and the Kondo lattice (see Section~\ref{Spin_polar} and references therein).

\textbf{vi.} In many local regions of the phase diagram, the RGM results agree with numerical simulations (see Section~\ref{Compare} and references therein).

\section{Results for a square lattice}
\label{Jall}

The basic ideas and the standard computational algorithm of the RGM were outlined above using a two-dimensional square lattice as an example. This section presents the main RGM results for three versions of the Heisenberg model on a square lattice — \(J_1\), \(J_1\)-\(J_2\), and \(J_1\)-\(J_2\)-\(J_3\) (with spin \(S = 1/2\) throughout this section).

The focus is placed primarily on results concerning the spin structure: the structure factor, short-range and long-range magnetic order, the spin excitation spectrum, and the evolution of all these quantities with temperature and exchange parameters.

Magnetization, magnetic susceptibility, and heat capacity are discussed only insofar as necessary to demonstrate the applicability of the method. Analytically tractable limits or, conversely, the fitting of numerical data to simple functional relationships are not systematized here. All this is presented in the cited literature; however, a thorough analysis of these topics would shift the focus of this review and significantly increase its length.

\subsection{\(J_1\) model}
\label{J1}

The most important result of the RGM for the model with the nearest-neighbor exchange \cite{Shimah91_JPSJ,Baraba92_JPSJ} is the very possibility of describing the spin structure in terms of a singlet for both the FM and AFM cases at temperatures \(T \geq 0\). In both cases, the average spin at each site is zero, and the constraint condition \(\langle \mathbf{\hat{S}}_{\mathbf{i}}\rangle = 0\), \(\langle \mathbf{\hat{S}}_{\mathbf{i}}^{2}\rangle = 3/4\) is exactly satisfied. The spin-spin correlator \(\langle S_{\mathbf{0}}^{\alpha }S_{\mathbf{r}}^{\alpha }\rangle\) is independent of \(\alpha\): \(\langle S_{\mathbf{0}}^{x}S_{\mathbf{r}}^{x}\rangle = \langle S_{\mathbf{0}}^{y}S_{\mathbf{r}}^{y}\rangle = \langle S_{\mathbf{0}}^{z}S_{\mathbf{r}}^{z}\rangle\), and the three branches of spin excitations are degenerate.

\textit{FM exchange \(J = -1\).}

At \(T = 0\), the structure factor \(c_{\mathbf{q}}\) exhibits a \(\delta\)-shaped peak at the zero point of the Brillouin zone \(\mathbf \Gamma = (0,0)\). The spin-spin correlator \(\langle S_{\mathbf{0}}^{\alpha }S_{\mathbf{r}}^{\alpha }\rangle\) is positive and approaches a constant value as \(\mathbf{r} \rightarrow \infty\) (the FM correlation length is infinite)
\begin{equation}
T = 0,\quad \langle S_{\mathbf{0}}^{\alpha }S_{\mathbf{r}}^{\alpha }
\rangle_{\mathbf{r\rightarrow \infty }}\rightarrow Const,\quad \alpha = x,y,z,
\label{Const_lim_cor}
\end{equation}
For the studied state with zero average spin, the constant in (\ref{Const_lim_cor}), the so-called effective magnetization \(m\), characterizes the spin structure
\begin{equation}
m=\left( \lim_{\mathbf{r\rightarrow \infty}}\,
|\langle \mathbf{S}_{\mathbf{0}}\mathbf{S}_{\mathbf{r}}\rangle |\right)^{1/2},
\label{Eff_m_FM}
\end{equation}
within the RGM for the FM case, one obtains \(m = 1/2\), as expected.

The spin excitation spectrum is close to the standard spin-wave spectrum (small differences depend on the choice of vertex corrections), and the gap is closed only at point \(\mathbf \Gamma\).

At \(T > 0\), the spectrum evolves, but no qualitative changes are observed. However, the height of the peak \(c_{\mathbf{q}}\) becomes finite, and its width now defines a finite correlation length. The spin-spin correlator decreases to zero asymptotics with increasing \(r\)
\begin{equation}
T > 0,\quad \langle S_{\mathbf{0}}^{\alpha }S_{\mathbf{r}}^{\alpha }
\rangle_{\mathbf{r\rightarrow \infty }}\rightarrow 0,\quad \alpha = x,y,z;
\end{equation}

Thus, the RGM provides a description of a singlet state with a finite correlation length and strong FM correlations.

\textit{AFM exchange \(J = +1\).}

At \(T = 0\), the \(\delta\)-shaped peak of the structure factor \(c_{\mathbf{q}}\) is located at the AFM point of the Brillouin zone \(\mathbf{Q} = (\pi,\pi)\). The spin-spin correlator alternates in sign (checkerboard pattern), and its modulus approaches a constant value as \(\mathbf{r} \rightarrow \infty\) (the AFM correlation length is infinite)
\begin{equation}
T = 0,\quad |\langle S_{\mathbf{0}}^{\alpha }S_{\mathbf{r}}^{\alpha }
\rangle|_{\mathbf{r\rightarrow \infty }}\rightarrow Const,\quad \alpha = x,y,z,
\label{Eff_m_AFM}
\end{equation}

The effective magnetization in (\ref{Eff_m_AFM}) for the AFM case is nonzero and equal to \(m = 0.3\), in agreement with the results obtained using many alternative analytical and numerical approaches, including two-sublattice approaches \cite{Manous91_RMP}. This agreement provides additional support for the validity of the RGM.

As in the FM case, the spin excitation spectrum is close to the standard spin-wave spectrum (minor differences depend on the choice of vertex corrections), with the gap closed at the zero point \(\mathbf{\Gamma} = (0,0)\) and at the AFM point \(\mathbf{Q} = (\pi,\pi)\).

At \(T > 0\), the peak height of \(c_{\mathbf{q}}\) becomes finite, the spin-spin correlator decreases to zero asymptotics with increasing \(r\), and the correlation length becomes finite
\begin{equation}
T > 0,\quad \langle S_{\mathbf{0}}^{\alpha }S_{\mathbf{r}}^{\alpha }
\rangle_{\mathbf{r\rightarrow \infty }}\rightarrow 0,\quad \alpha = x,y,z;
\end{equation}

However, unlike the FM case, the spectrum undergoes a qualitative change at finite temperature. Specifically, the gap opens and increases with increasing temperature at the AFM point \(\mathbf{Q} = (\pi,\pi)\).

This is a key difference between the RGM and the spin-wave and related approaches. The spherically symmetric approach makes it possible to describe the singlet state of a system with strong AFM correlations.

The mathematical reason for this is as follows: translational symmetry is not broken in the RGM, and no lattice doubling occurs in the AFM case. The analysis is carried out in the full (not reduced magnetic) Brillouin zone, where the points \(\mathbf{Q} = (\pi,\pi)\) and \(\mathbf{\Gamma} = (0,0)\) are inequivalent.

\subsection{\(J_1\)-\(J_2\) model}
\label{J1-J2}

The \(J_1\)-\(J_2\) model has been extensively studied in the literature \cite{Baraba94_JPSJ,Baraba94_JETP_R,Siurak01_PRB,Mikhey06_PLA,Baraba07_PLA,Mikhey09_PLA, Hartel10_PRB,Hartel13_PRB,Mikhey13_JL_R,Mikhey15_JETP_R,Baraba15_JL_R,Mikhey16_JMMM} (early reviews of some ideas in this area can be found in \cite{Richte04_BookChap,Froebri06_PR}).

The initial motivation for investigating this model was, of course, related to HTSC cuprates, in which the spin subsystem is quasi-two-dimensional and possesses square symmetry. More than fifteen compounds have been identified in which experiments indicate frustrated \(J_1\)-\(J_2\) magnetism \cite{Feldke95_PRB,Feldke98_PRB,Melzi00_PRL,Melzi01_PRB,Rosner03_PRB,Manaka03_PRB,Kaul04_JMMM, Kageya05_JPSJ,Kasina06_PRB,Skoula07_JMMM,Nath08_PRB,Carret09_PRB,Tsirli09_PRB,Tsirli09_PRBa,Skoula09_EL, Tsirli10_PRB,Tsirli11_PRB,Guchha22_PRB}. In addition to the HTSC cuprates, these include, for example, Pb\(_2\)VO(PO\(_4\))\(_2\), (CuCl)LaNb\(_2\)O\(_7\), SrZnVO(PO\(_4\))\(_2\), BaCdVO(PO\(_4\))\(_2\), K\(_2\)CuF\(_4\), Cs\(_2\)CuF\(_4\), Cs\(_2\)AgF\(_4\), La\(_2\)BaCuO\(_5\), Rb\(_2\)CrCl\(_4\), and others.

These compounds fill the entire upper half of the \(J_1\)-\(J_2\) circle \cite{Schmid07_JPCM,Mikhey13_JL_R,Mikhey15_JETP_R,Mikhey16_JMMM}. Moreover, many of them are characterized by frustration leading to suppression of long-range order \cite{Kaul04_JMMM,Kageya05_JPSJ,Skoula07_JMMM,Nath08_PRB,Carret09_PRB,Skoula09_EL, Tsirli09_PRB,Tsirli09_PRBa}.

Two further clarifications are necessary.

First. Early work on the \(J_1\)-\(J_2\) model was motivated by the idea of reproducing the magnetic properties of HTSC cuprates. Some of these attempts were made to describe the experimentally observed `subtle properties' of the spin spectrum of HTSC cuprates, such as the so-called `resonance mode' or `hour-glass spectrum' (see \cite{Plakida10_Book_Theor} for details). To this end, additional parameters and considerations were introduced into the standard scheme. We will not go into such details here, and below we will present the basic, fundamental properties of the spin spectra in the \(J_1\)-\(J_2\) model\footnote{Most of the quantitative results in this section correspond to the standard RGM implementation in the one-vertex approximation.}.

Second. For the same reason — the description of HTSC — many studies considered theoretical models containing both a magnetic background and charge carriers (interacting with it). This area is beyond the scope of this review and will be only briefly mentioned in Section \ref{Spin_polar}.

Before proceeding to a systematic description of the possible spin structures in the \(J_1\)-\(J_2\) model, several general introductory remarks are necessary.

At \(T = 0\), the structure factor \(c_\mathbf{q}\) can either contain a \(\delta\)-shaped contribution or be finite throughout the Brillouin zone. In the former case, this is a phase with long-range order, the structure of which is determined by the peak position. In the latter case, it is a spin-liquid phase without long-range order, with the short-range order being determined by the structure factor profile.

At \(T \neq 0\), long-range order is impossible in the RGM under any exchange conditions, and \(c_\mathbf{q}\) is finite everywhere. The point \(\mathbf{q}_0\) — the position of the maximum of the structure factor in the Brillouin zone — determines the structure of the short-range order, and the half-width of the \(c_\mathbf{q}\) peak defines the correlation length relative to the corresponding order (ignoring equivalent points in the full Brillouin zone). Thus, the point \(\mathbf{q}_0\) is the quantum analogue of the control point in the classical model.

\subsubsection{Spin order at zero temperature}

\begin{figure}[h]
  \centering
  \includegraphics[width=0.45\textwidth]{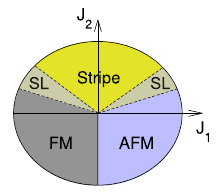}
  \caption{\small (Color online) Schematic of the quantum \(S = 1/2\) \(J_1\)-\(J_2\) model at \(T = 0\). A disordered spin-liquid (SL) phase is realized between the AFM and stripe phases with long-range order. A second spin-liquid region appears between the FM and stripe long-range orders. The spin-liquid boundaries are shown tentatively, without exact correspondence to the calculated data.}
  \label{fig_Circle_Quant}
\end{figure}

At \(T = 0\), depending on the ratio of exchange interactions, five distinct spin structures are realized: three phases with long-range order (AFM, stripe, and FM) and two spin-liquid regions (see Fig.~\ref{fig_Circle_Quant}). As one moves along the \(J_1\)-\(J_2\) circle, a cascade of quantum phase transitions occurs.

\begin{figure}[h]
  \centering
  \includegraphics[width=0.85\textwidth]{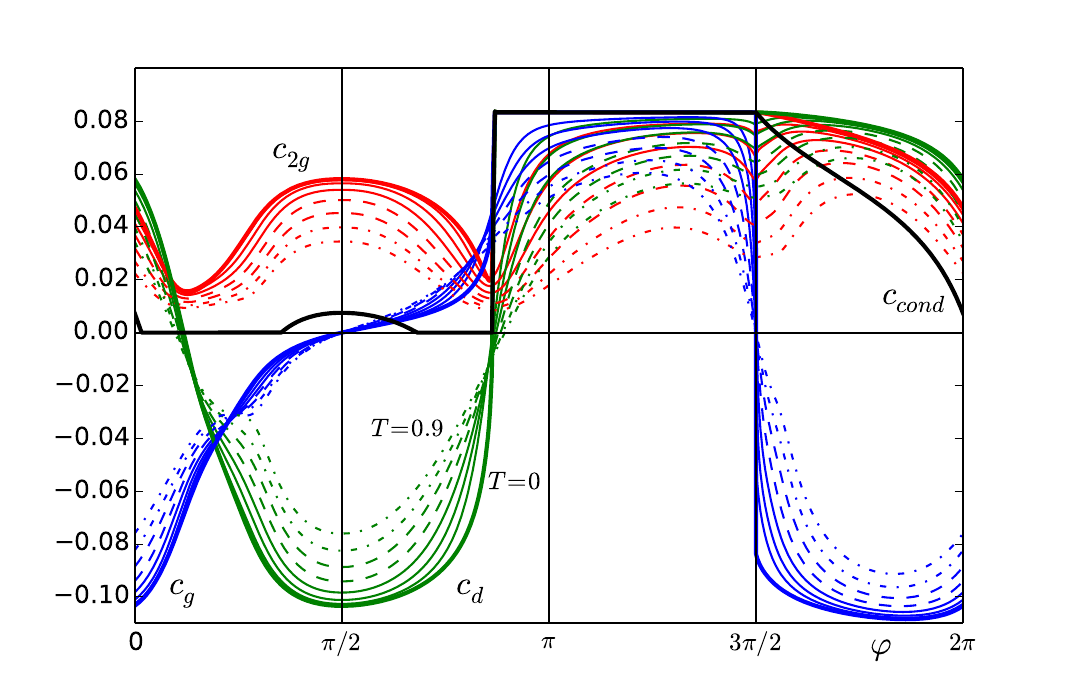}
  \caption{\small (Color online) \(J_1\)-\(J_2\) model. Spin-spin correlators at the nearest neighbors (\(c_g\), blue), second nearest neighbors (\(c_d\), green), and third nearest neighbors (\(c_{2g}\), red) as functions of parametric angle \(\varphi\) (\(J_{1} = \cos \varphi\), \(J_{2} = \sin \varphi\)). The thick solid lines correspond to temperature \(T = 0\), the thin solid lines to \(T = 0.3, 0.4, 0.5\), the dashed lines to \(T = 0.6, 0.7\), and the dotted lines to \(T = 0.8, 0.9\). The spin condensate \(c_{cond}\) (on a different scale) is shown by the solid black line. Data from \protect\cite{Mikhey16_JMMM}.}
  \label{fig_J12_Corrs}
\end{figure}

\begin{figure}[h]
  \centering
  \includegraphics[width=0.85\textwidth]{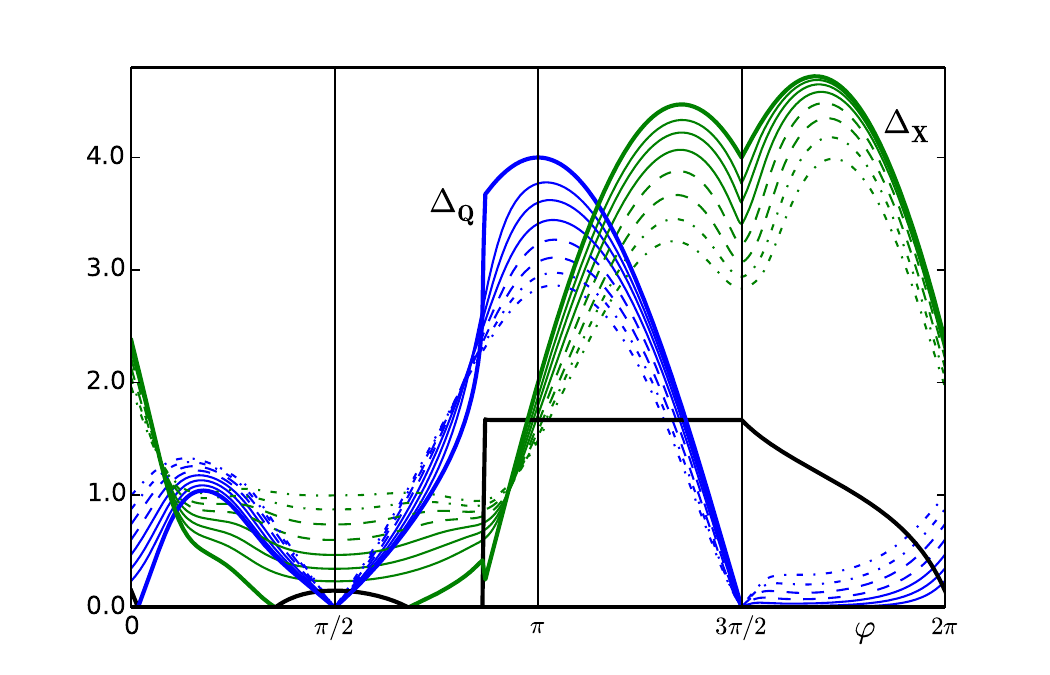}
  \caption{\small (Color online) Gaps in the spin excitation spectrum \(\Delta_{\mathbf{Q}}\) (blue) and \(\Delta_{\mathbf{X}}\) (green) at symmetric points of the Brillouin zone \(\mathbf{Q} = (\pi,\pi)\) and \(\mathbf{X} = (0,\pi), (\pi,0)\) as functions of parametric angle \(\varphi\) (\(J_{1} = \cos \varphi\), \(J_{2} = \sin \varphi\)). The thick solid lines correspond to temperature \(T = 0\), the thin solid lines to \(T = 0.3, 0.4, 0.5\), the dashed lines to \(T = 0.6, 0.7\), and the dotted lines to \(T = 0.8, 0.9\). The spin condensate \(c_{cond}\) (on a different scale) is shown by the solid black line. Data from \protect\cite{Mikhey16_JMMM}.}
  \label{fig_J12_Gaps}
\end{figure}

The evolution of spin-spin correlators over the entire \(J_1\)-\(J_2\) circle is shown in Fig.~\ref{fig_J12_Corrs}. Gaps in the spin excitation spectrum at symmetric points of the Brillouin zone are presented in Fig.~\ref{fig_J12_Gaps}. The trigonometric parameterization of the exchanges introduced in Section \ref{Class_phases} is used: \(J_{1} = \cos \varphi\), \(J_{2} = \sin \varphi\), \(0 \leq \varphi \leq 2\pi\), \(\sqrt{J_{1}^{2}+J_{2}^{2}} = 1\)\footnote{All energy quantities in this section (except Fig.~\ref{fig_Spectrum_p}) are normalized to \(\sqrt{J_{1}^{2}+J_{2}^{2}}\).}.

\textbf{Upper half of the \(J_1\)-\(J_2\) circle.}

\(0 \to SL_1\). On the upper semicircle, between \(\varphi = 0\) and the lower boundary of the spin-liquid phase \(SL_1\), the \textit{AFM phase with long-range order} is realized. The structure factor \(c_{\mathbf{q}}\) has a \(\delta\)-shaped peak at the AFM point of the Brillouin zone \(\mathbf{Q} = (\pi,\pi)\). The gap in the spin excitation spectrum at point \(\mathbf{Q}\) is closed\footnote{As in the \(J_1\) model, in \(J_1\)-\(J_2\) the gap at the zero point \(\mathbf{\Gamma }=(0,0)\) is closed at any temperature and any exchanges.}. The modulus of the spin-spin correlator at infinity approaches a constant value, and its sign changes in a checkerboard pattern
\begin{equation}
\langle S_{\mathbf{0}}^{\alpha }S_{\mathbf{r}}^{\alpha }\rangle^{\mathrm{afm}}
_{\mathbf{r\rightarrow \infty }} = c^{\mathrm{afm}}_{\mathrm{cond}}(-1)^{n_{x}+n_{y}},\quad
\alpha =x,y,z, \quad
\mathbf{r} = n_{x}\mathbf{g}_{x}+n_{y}\mathbf{g}_{y}
\label{afm_cond}
\end{equation}
As \(\varphi\) increases, the spin condensate \(c^{\mathrm{afm}}_{\mathrm{cond}}\) decreases, and a quantum phase transition to a \textit{spin liquid without long-range order} occurs at point \(SL_1\).

\(SL_1 \to SL_2\). In the spin liquid, the peak in the structure factor at point \(\mathbf{Q}\) becomes finite, with its width determining the correlation length. The gap in the spectrum at point \(\mathbf{Q}\) is open, but small. Short-range order retains the AFM structure (see Fig.~\ref{fig_J12_Corrs}); the nearest-neighbor correlator \(c_g < 0\), the next-nearest and third-neighbor correlators \(c_d > 0\), \(c_{2g} > 0\).

With increasing \(\varphi\) when moving from \(SL_1\) to the upper boundary of the spin liquid \(SL_2\), the gaps decrease, and the maxima of the structure factor at the stripe points of the Brillouin zone \(\mathbf{X} = (0,\pi), (\pi,0)\) increase. At the AFM point, on the contrary, the gap increases, and the peak of \(c_{\mathbf{q}}\) decreases. Thus, short-range order evolves towards the stripe structure. At the upper boundary of the spin liquid \(SL_2\), a quantum phase transition to the \textit{stripe phase with long-range order} occurs.

\(SL_2 \to SL_3\). This is the stripe phase region. The structure factor has a \(\delta\)-shaped peak at the stripe points of the Brillouin zone \(\mathbf{X} = (0,\pi), (\pi,0)\). The gap in the spectrum of spin excitations at the points \(\mathbf{X}\) is closed. The modulus of the spin-spin correlator at infinity approaches a constant value. In the RGM, the stripe phase is a quantum superposition of horizontal and vertical stripes, and the correlator at infinity has the form
\begin{equation}
\langle S_{\mathbf{0}}^{\alpha }S_{\mathbf{r}}^{\alpha }
\rangle_{\mathbf{r\rightarrow \infty}}^{\mathrm{stripe}} =
c_{\mathrm{cond}}^{\mathrm{stripe}} \frac{1}{2}[(-1)^{n_{x}}+(-1)^{n_{y}}]
\quad
\alpha =x,y,z, \quad
\mathbf{r} = n_{x}\mathbf{g}_{x}+n_{y}\mathbf{g}_{y}
\label{stripe_cond}
\end{equation}

The point \(\varphi = \pi/2\) of the \(J_1\)-\(J_2\) circle is highlighted. It corresponds to \(J_1 = 0\), \(J_2 = 1\). From geometric considerations, it is obvious that for such exchanges, the lattice splits into two diagonally shifted, noninteracting sublattices. This leads, for example, to a divergence of the magnetic susceptibility. An arbitrarily small value of \(J_1 \neq 0\) eliminates the divergence\footnote{\label{foot_J10}Moreover, at the points \(\varphi = \pi/2\) and the opposite point \(\varphi = 3\pi/2\), the gap \(\Delta_{\mathbf{Q}}\) is closed at any \(T\) (see Fig.~\ref{fig_J12_Gaps}). Formally, this follows from the analytical expression for \(\Delta_{\mathbf{Q}}\). The physical explanation is that the system of two noninteracting sublattices is degenerate with respect to their mutual rotation. Therefore, transferring a spin excitation to a neighboring site requires no energy.}.

As \(\varphi\) increases, the magnitude of the spin condensate \(c^{\mathrm{cond}}_{\mathrm{stripe}}\) decreases, and the next quantum phase transition occurs at point \(SL_3\) — to the \textit{second spin-liquid region without long-range order}.

\(SL_3 \to SL_4\). In the second spin liquid region, the peaks of the structure factor at the stripe points of the Brillouin zone become finite, as does the correlation length. The gaps at the \(\mathbf{X}\) points are open. Short-range order retains the stripe structure (see Fig.~\ref{fig_J12_Corrs}), characterized by \(c_d < 0\), \(c_{2g} > 0\); the sign of \(c_g\) can change.

With further increase in \(\varphi\) when moving from \(SL_3\) to \(SL_4\), the maximum of the structure factor at the zero point of the Brillouin zone \(\mathbf{\Gamma }=(0,0)\) increases (the gap in the spectrum at this point is always closed). Short-range order evolves toward the FM structure.

At point \(SL_4\), another quantum phase transition occurs — to the \textit{FM region with long-range order}.

\(SL_4 \to \pi\). FM region. The \(\delta\)-shaped peak of the structure factor at the zero point of the Brillouin zone \(\mathbf{\Gamma }=(0,0)\) (the gap is also closed there). All spin-spin correlators are positive. The correlator at infinity approaches a constant sign \(c^{\mathrm{cond}}_{\mathrm{fm}}\).

\textbf{Lower half of the \(J_1\)-\(J_2\) circle.}

The lower semicircle is of no experimental interest — compounds with \(J_1 < 0\), \(J_2 < 0\) (third quadrant, \(\pi < \varphi < 3 \pi /2\)) and with \(J_1 > 0\), \(J_2 < 0\) (fourth quadrant, \(3 \pi /2 < \varphi < 2 \pi\)) have not been detected.

From a theoretical perspective, this region is also not very interesting due to the absence of frustration: in the third quadrant, both exchanges favor the FM phase, while in the fourth quadrant, they favor the AFM phase. The only noteworthy feature is the transition at \(\varphi = 3 \pi /2\), when the correlator at the nearest neighbors changes its sign in a jump-like manner.

\subsubsection{Spin structure at finite temperature}

At \(T > 0\), long-range order is absent in the RGM for any exchanges. Spin-spin correlators at infinity are zero. Gaps in the spectrum at symmetric points of the Brillouin zone are open (except for the point \(\mathbf{\Gamma} = (0,0)\), where the gap is always closed), and the peaks of the structure factor have finite height.

However, short-range order retains the structure of the corresponding zero-temperature phase up to high temperatures \(T \sim 1\). For example, for the region lying above the AFM phase, short-range order still preserves a checkerboard pattern (\(c_g < 0\), \(c_d > 0\), \(c_{2g} > 0\)), see Fig.~\ref{fig_J12_Corrs}.

The appearance of the spin excitation spectrum also resembles that at \(T = 0\), with the difference that the gaps at the corresponding symmetric points of the Brillouin zone are now nonzero.

\begin{figure}[h]
\begin{center}
\begin{minipage}{17pc}
\includegraphics[width=17pc, trim=2pc 0pc 0pc 0]{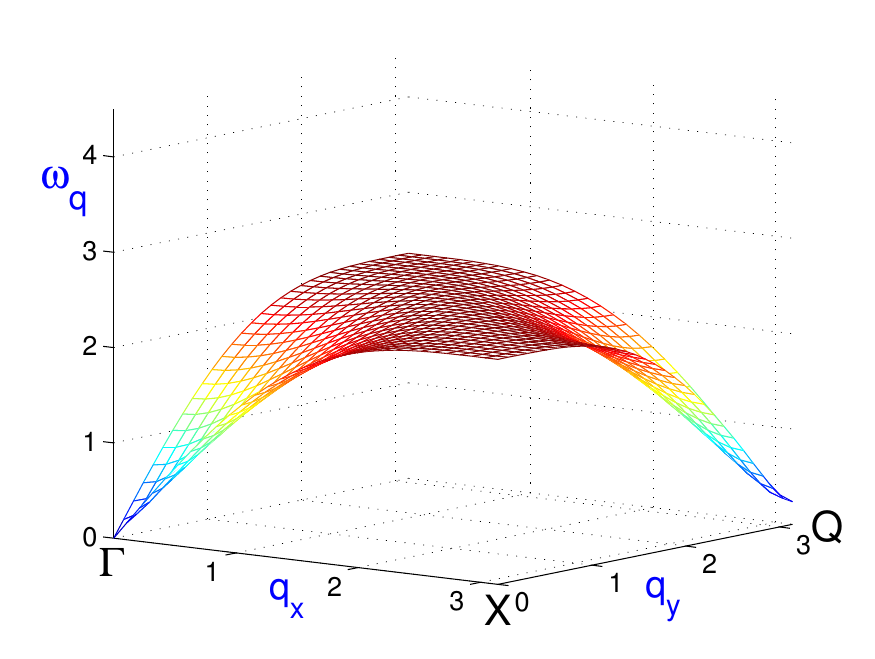} \\ \hspace*{3cm} {\textbf{a}}
\end{minipage}\hspace{.5pc}
\begin{minipage}{17pc}
\includegraphics[width=17pc, trim=2pc 0pc 0pc 0]{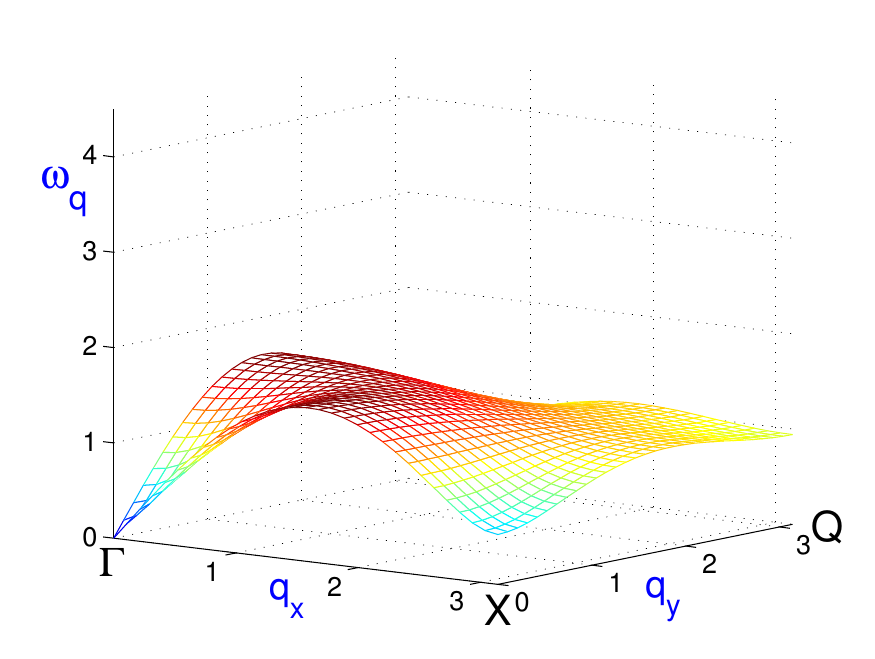} \\ \hspace*{3cm} {\textbf{b}}
\end{minipage}\hspace{.5pc}
\begin{minipage}{17pc}
\includegraphics[width=17pc, trim=2pc 0pc 0pc 0]{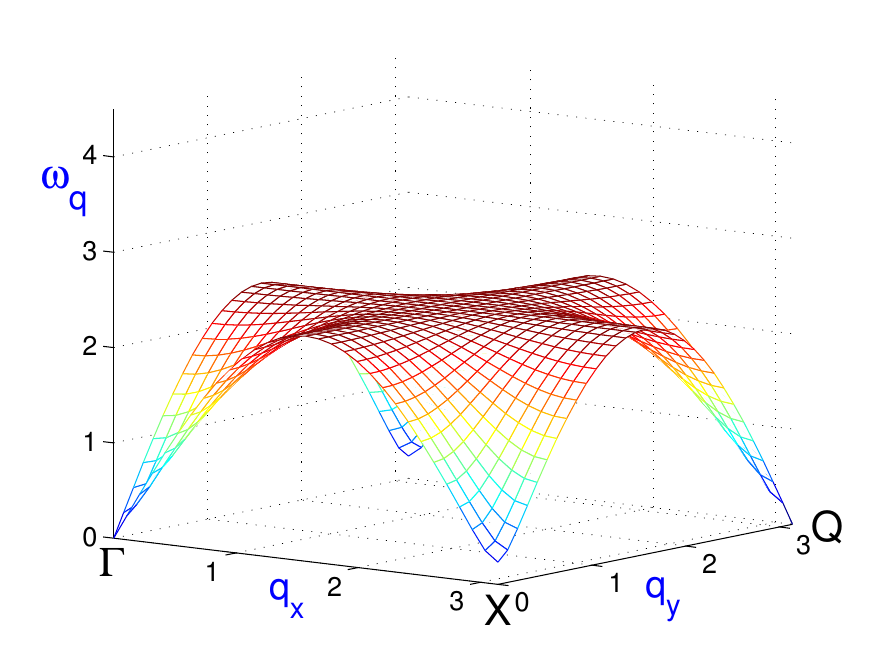} \\ \hspace*{3cm} {\textbf{c}}
\end{minipage}
\begin{minipage}{17pc}
\includegraphics[width=17pc, trim=2pc 0pc 0pc 0]{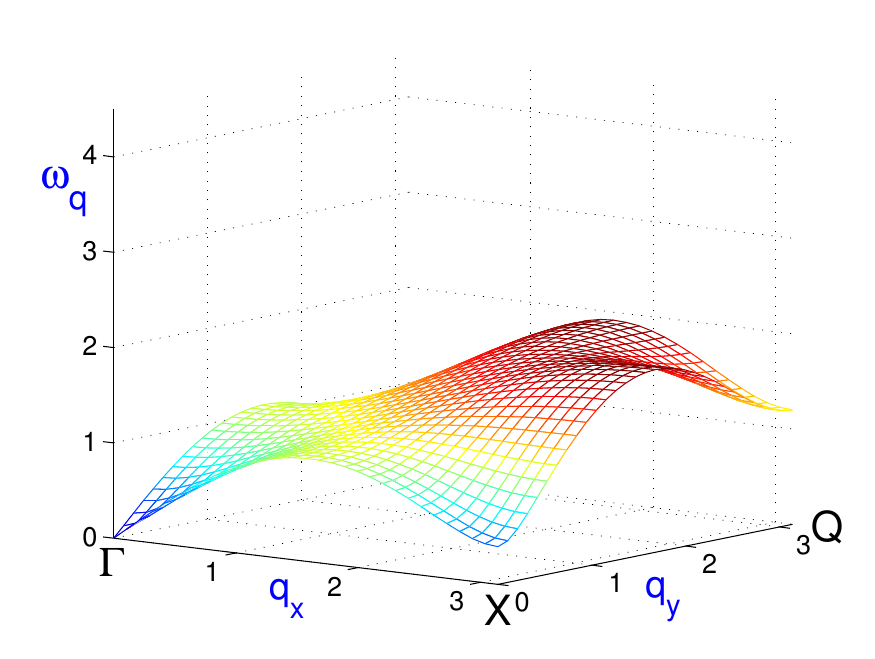} \\ \hspace*{3cm} {\textbf{d}}
\end{minipage}
\caption{\small (Color online) Spin excitation spectra at \(T=0.3\) in the upper half of the \(J_1\)-\(J_2\) circle. Graph {\textbf{a}}: \(\varphi=0\), corresponds to the AFM zero-temperature phase, the gap \(\Delta_{\mathbf{Q}}\) is small. {\textbf{b}}: \(\varphi=\pi/4\) — above the spin-liquid region, the gaps at symmetric points are comparable. {\textbf{c}}: \(\varphi=\pi/2\) — corresponds to the zero-temperature stripe phase, the gaps \(\Delta_{\mathbf{X}}\) are small (for \(\Delta_{\mathbf{Q}}\) at \(\varphi = \pi/2\) see text). {\textbf{d}}: \(\varphi=3\pi/4\) — also above the spin-liquid region, the gaps \(\Delta_{\mathbf{X}}\) are somewhat larger than in the previous graph. The gap at the zero point \(\Delta_{\mathbf{\Gamma}}\) is closed for any exchanges and temperature. A quarter of the Brillouin zone is shown (\protect\cite{Mikhey15_JETP_R}).}
\label{fig_Specta_J12}
\end{center}
\end{figure}

Figure~\ref{fig_Specta_J12} shows examples of the spin spectrum at \(T = 0.3\) for a wide range of exchanges on the upper semicircle. Graph \textbf{a} corresponds to \(\varphi = 0\), the region of the zero-temperature AFM phase with long-range order; here, the gap at the AFM point \(\Delta_{\mathbf{Q}}\) is small (recall that the gap at the zero point \(\Delta_{\mathbf{\Gamma}}\) is closed for any exchanges and temperature). Graph \textbf{b}, \(\varphi = \pi/4\) — above the spin-liquid region; the gaps at the symmetric points are comparable. Graph \textbf{c}, \(\varphi = \pi/2\) — above the stripe phase with long-range order; the gaps at the stripe points \(\Delta_{\mathbf{X}}\) are small (the gap \(\Delta_{\mathbf{Q}}\) is always zero for the exact equality \(\varphi = \pi/2\), see footnote \ref{foot_J10}). Graph \textbf{d}, \(\varphi = 3\pi/4\) — also above the spin-liquid region; the gaps at the points \(\Delta_{\mathbf{X}}\) are somewhat larger than in the previous graph.

\begin{figure}[h]
  \centering
  \includegraphics[width=0.75\textwidth]{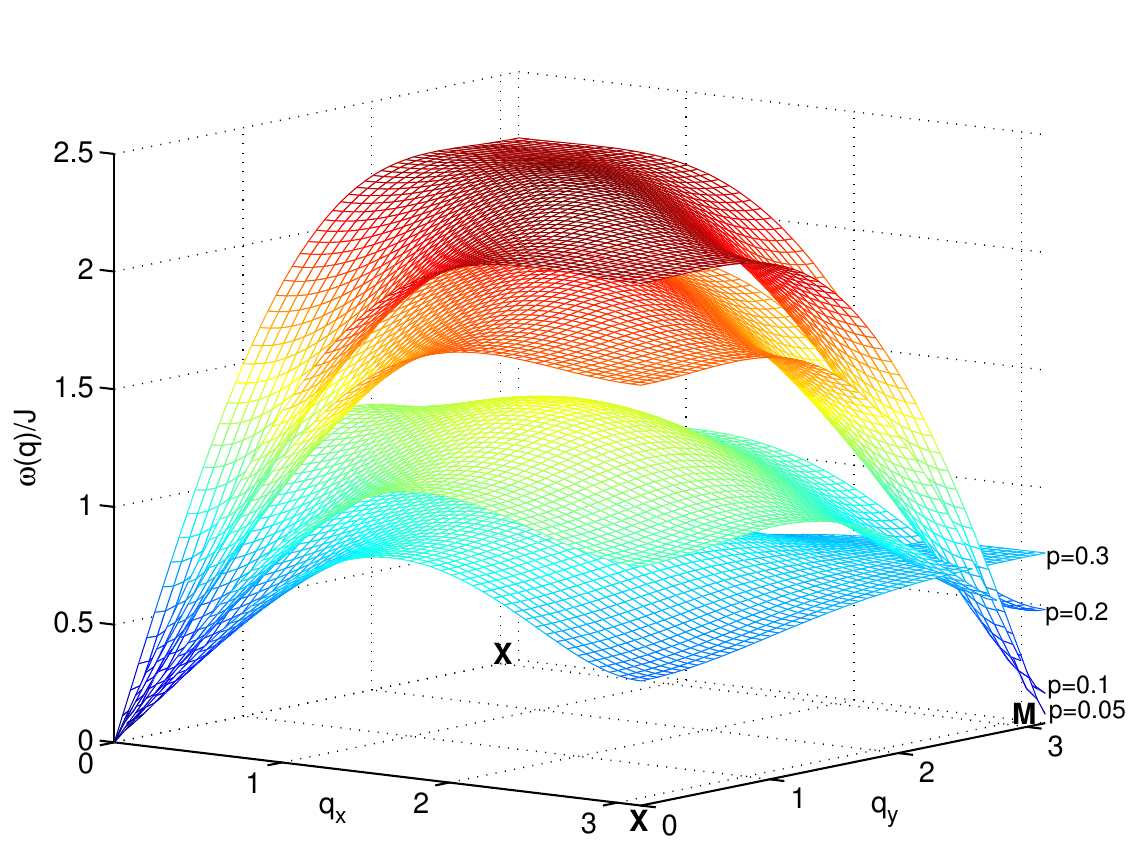}
  \caption{\small (Color online) Evolution of the spin excitation spectrum at a finite temperature in the first quadrant of the \(J_1\)-\(J_2\) circle with increasing \(\varphi\) from \(\varphi = 0\). In this figure from \protect\cite{Baraba11_TMP_R}, the standard frustration parameter for the first quadrant \(p=J_2/J\), \(J = J_1+J_2\) is used (an increase in \(p\) corresponds to an increase in \(\varphi\)), and an alternative notation for the AFM point \(\mathbf{M} \equiv \mathbf{Q} = (\pi,\pi)\). It is seen that with increasing \(p\) (i.e., with increasing \(\varphi\)), the gap at the AFM point increases, and the gaps at the stripe points \(\mathbf{X}\) decrease. The short-range order transforms from AFM-like to stripe-like. Here the temperature is \(T = 0.1\) in units of \(J = J_1+J_2\), and the spectrum is in the same units.}
  \label{fig_Spectrum_p}
\end{figure}

An example of the evolution of the spectrum in a local region — as \(\varphi\) increases from \(\varphi = 0\) — is shown in Fig.~\ref{fig_Spectrum_p}. It can be seen that the gap at the AFM point increases while the gaps at the stripe points decrease (although both remain nonzero); i.e., the spectrum transforms from AFM-like to stripe-like. Simultaneously, the (finite) maximum of the structure factor decreases at the AFM point and increases at the stripe point.

Regarding the lower semicircle at \(T > 0\), as noted above, frustration is absent; both exchanges operate in the same direction. This leads to `antifrustration' — in the third sector \(J_1 < 0, J_2 < 0\), a `superferromagnet' emerges, and in the fourth sector \(J_1 > 0, J_2 < 0\), a `superantiferromagnet' is realized. In these regions, at \(T > 0\) the correlation length is exponentially large up to temperatures \(T \sim 1\), i.e., on the order of \(J = \sqrt{J_{1}^{2}+J_{2}^{2}}\).

\textit{Susceptibility and heat capacity.} As promised at the beginning of this section, we present mainly results concerning the spin structure. However, we will mention some other quantities calculated in the RGM.

A significant portion of the studies cited above also determined the behavior of the spin susceptibility (usually static) and heat capacity.

At \(T = const\), the dependence of the static susceptibility \(\chi(\mathbf{q},\omega = 0)\) on the exchange parameters is expected: on the \(J_1\)-\(J_2\) circle, \(\chi(\mathbf{q},\omega = 0)\) calculated at a specific symmetric point \(\mathbf{q}_0\) of the Brillouin zone exhibits a maximum in the region of the corresponding spin structure. The temperature behavior of the susceptibility \(\chi (\mathbf{q},\omega,T)\), at least for the static one, shows a greater variety of regimes depending on the exchange parameters. Some studies have also analyzed the `local spin susceptibility' \(\chi _{2D}(\omega ,T)=\int d\mathbf{q\,}\mathrm{Im} \chi (\mathbf{q},\omega,T)\), which has a scaling property (see more details at the end of Section \ref{Renorm}). We will avoid a full systematization of these results.

The heat capacity \(C_V\) displays a simpler behavior on the \(J_1\)-\(J_2\) circle. For any exchanges, \(C_V\) has a standard temperature dependence: \(C_V \to 0\) as \(T \to 0\) and a single maximum at a finite temperature, the position of which depends on the exchanges. An exception occurs in the region near the stripe \(\leftrightarrow\) FM transition, where an additional low-temperature maximum appears (first observed in \cite{Hartel10_PRB}). In the literature, it is commonly attributed to frustration.

At any fixed temperature, \(C_V\) exhibits local minima in regions above ordered phases and in regions above transitions between phases. Both cases correspond to local extrema of the energy at \(T = Const\); the first — to maxima, the second — to minima \cite{Mikhey16_JMMM}.

\subsubsection{Comparison with experiment and with numerical methods}
\label{Compare}

In a significant portion of studies on the \(J_1\)-\(J_2\) model that used the spherically symmetric approach (on variations of the method, see Section \ref{Relat}), comparisons of the results with numerical methods were carried out. Without delving into numerous details, it can be stated that the agreement with numerical results is at least satisfactory, and usually good. For completeness, related models in the spherically symmetric approach are also mentioned below.

Comparisons with numerical results were carried out not only for variants of the Heisenberg model on a square lattice \cite{Antsyg08_PRB,Junger05_PRB,Junger08_PRB,Darrad08_PRB,Junger04_PRB}, but also for a triangular lattice \cite{Rubin05_PLA,Antsyg08_PRB,Rubin17_JPCS}, the Kagome lattice, and pyrochlore \cite{Mueller18_PRB,Mueller17_PRB}. Numerous comparisons with variants of a one-dimensional chain \cite{Hartel08_PRB,Richte09_JPCS,Hartel11_PRB,Hutak22_EPJB,Hutak23_EPJB,Antsyg08_PRB,Junger04_PRB} have been made. Finally, a similar procedure was carried out for the \(t-J\) model — the most common model with a magnetic background and charge carriers (see Section \ref{Spin_polar}) \cite{Siurak01_PRB,Sherma02_PRB,Vladim09_PRB}.

Regarding the numerical methods used, the most common is, of course, exact diagonalization, typically using the Lanczos algorithm \cite{Hartel08_PRB,Richte09_JPCS,Hartel11_PRB,Hutak22_EPJB,Hutak23_EPJB, Rubin05_PLA,Antsyg08_PRB,Rubin17_JPCS,Junger05_PRB,Siurak01_PRB, Junger04_PRB,Sherma02_PRB,Vladim09_PRB}. However, other methods were also used: quantum Monte Carlo \cite{Antsyg08_PRB,Junger08_PRB,Sherma02_PRB}, the coupled cluster approach \cite{Darrad08_PRB}, and high-temperature expansion \cite{Hutak23_EPJB,Mueller18_PRB,Mueller17_PRB}. For the basic one-dimensional problem, the natural benchmark is the Bethe ansatz \cite{Antsyg08_PRB,Junger04_PRB}.

In summary, it can be said that the available numerical methods confirm the complete adequacy of the RGM and related approaches.

\textit{Comparison with experiment.} In the early stages of theoretical development, motivated by the emergence of HTSC cuprates, comparison with experiment was essentially a mandatory requirement for papers on the topic. Almost every paper compared the main characteristics of the spin excitation spectrum, fine details of the spectrum (e.g., the `hour-glass spectrum'), the correlation length, the position of the structure factor maximum, the spin susceptibility, and other magnetic and thermodynamic characteristics with experiment. A detailed analysis of this data set, especially in the context of several different HTSC theories, is a formidable task, far beyond the scope of this review, and will not be presented here.

\subsection{\(J_1\)-\(J_2\)-\(J_3\) model}
\label{J1-J2-J3}

We now turn to the \(J_1\)-\(J_2\)-\(J_3\) model.

The Hamiltonian has the form (\ref{Hamilt}) with three exchanges. Recall that for a square lattice, \(J_1\) is the exchange along the side of the square, \(J_2\) is along the diagonal, and \(J_3\) is along twice the side (see Fig.~\ref{fig_Neighbours}).

The calculation scheme remains the same as that described in Section \ref{calc_scheme}, although the expressions for the Green's function become quite cumbersome (provided in the Appendix).

Below, the `spherical' parameterization of exchanges mentioned in Section \ref{Class_phases} is used
\begin{eqnarray}
J_{1} &=&\cos (\psi )\cos (\varphi ),\quad J_{2}=\cos (\psi )\sin (\varphi
),\quad J_{3}=\sin (\psi ), \notag \\
0 &\leq &\varphi \leq 2\pi ,\quad -\pi/2\leq \psi \leq \pi/2,\quad \sqrt{%
J_{1}^{2}+J_{2}^{2}+J_{3}^{2}}=1
\label{Param_123bis}
\end{eqnarray}

This section considers the case of low but nonzero temperatures. This is sufficient to detect all types of emerging magnetic structures.

\begin{figure}[h]
  \centering
  \includegraphics[width=0.75\textwidth]{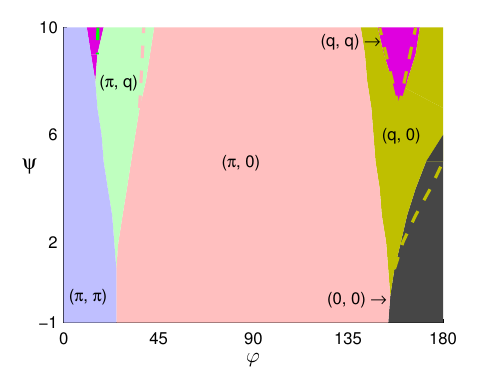}
  \caption{\small (Color online) Regions of the phase plane corresponding to different short-range order structures (axis angles in degrees). The positions of the maximum of the structure factor are indicated. \((\pi,\pi)\) — AFM, \((\pi,0)\) — stripe, \((0,0)\) — FM, \((\pi,q)\), \((q,0)\) and \((q,q)\) — three types of helicoids. Solid boundaries correspond to \(T = 0.4\), dashed boundaries to \(T = 0.2\). At lower temperatures, the boundaries stabilize. Data from \protect\cite{Mikhee18_JETP_R,Valiul19_JPCM}.}
  \label{fig_PD_J123_quant}
\end{figure}

Figure~\ref{fig_PD_J123_quant} shows a section of the most interesting, fully frustrated quarter of the phase diagram, indicating the various short-range order structures. The following types of short-range order are realized: AFM, stripe, FM, and three types of quantum helices. As in the classical limit (see expressions (\ref{Class_Ener_123})), they correspond to the positions of the control point (maximum of the structure factor), respectively: \(\mathbf{\Gamma} = (0,0)\), \(\mathbf{X} = (0,\pi), (\pi,0)\), \(\mathbf{Q} = (\pi,\pi)\); for helices — to its motion along one side or the other of the Brillouin zone, as well as along its diagonal (helicoids \((0,q)\), \((q,0)\), and \((q,q)\)).

Figure~\ref{fig_PD_J123_quant} corresponds to finite temperatures, so only short-range order is detected everywhere. The picture remains qualitatively unchanged down to extremely low temperatures. At \(T = 0\), regions of long-range order arise, separated by a spin-liquid phase. The issue of spin-liquid boundaries has been studied in detail for \(J_3 = 0\) (see references in the previous section) and remains largely unexplored for \(J_3 \neq 0\), in particular for \(J_3 > 0\).

\subsubsection{Frustrated helices}
\label{frust_helix}

The most important new property that arises with the addition of \(J_3\) is the possibility of helical spin structures, both commensurate and incommensurate. This effect is nontrivial, since helices (see, e.g., \cite{Boriso20_UFN_R}), including those in 2D, are usually associated with the Dzyaloshinskii-Moriya interaction (DMI, \cite{Dzyalo57_JETP_R,Dzyalo58_JPCS,Moriya60_PRL,Moriya60_PR}; references to recent works can be found in \cite{Nocula23_PRB,Silva24_PRE}; for a general classification of helices, see \cite{Izyumov84_UFN_R}). For the realization of a DMI helix, in particular, the breaking of symmetry with respect to inversion is necessary.

In the \(J_1\)-\(J_2\)-\(J_3\) model, helices arise without breaking this symmetry, without DMI interaction. The reason for their appearance here is frustration. Therefore, such helices are sometimes called frustrated, and in DMI, they are called nonfrustrated \cite{Soroki19_JMMM}.

Frustrated helices — in the classical limit \(S \gg 1\) — have been known since the middle of the last century. Here, the ultraquantum case \(S = 1/2\) is considered in the RGM interpretation.

It is important to keep the following in mind. Quantum helical states in the RGM should not be thought of in a semiclassical manner as a lattice of spin arrows changing direction from site to site. Any spin state, including a helix, is singlet, with the average spin at a site equal to zero. The helix structure can be identified in one of three ways: by the pattern of correlators, by the position of the structure factor maximum, or by features of the spin excitation spectrum.

The first method seems obvious, but it is less intuitive. It is necessary to construct a picture of the correlators for the classical analogue of the desired quantum helix and compare it with the correlators in the quantum case (examples in \cite{Mikhee18_JETP_R}). This procedure is complicated by two circumstances. First, the correlation length at \(T > 0\) is finite, and the correlators decay with distance. Second, since none of the initial symmetries are broken in the RGM, the helical state is a superposition of quantum helices defined by equivalent points in the full Brillouin zone.

\begin{figure}[h]
  \centering
  \includegraphics[width=0.75\textwidth]{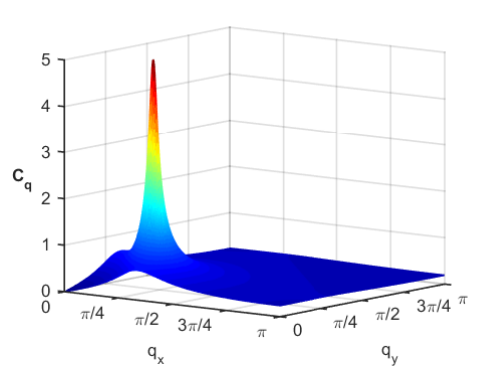}
  \caption{\small (Color online) Example of a structure factor for a helicoid. The sharp maximum \(c_\mathbf{q}\) on the diagonal of the Brillouin zone (helicoid \((q,q)\)) at low temperature indicates the nature of the short-range order. The peak width determines the correlation length. \(T=0.02\), \(\varphi=160^{\circ}\), \(\psi=10^{\circ}\). Data from \protect\cite{Mikhee18_JETP_R}.}
  \label{fig_Cq_Diag}
\end{figure}

The second and third criteria (already mentioned in the previous section) are closely related. At not too high temperatures, the structure factor has a local maximum at the control point of the Brillouin zone, which determines the structure of the helix (and equivalents). At the same point, the spin excitation spectrum has a local minimum. At \(T = 0\) and in the presence of long-range order, the structure factor maximum is \(\delta\)-shaped, and the spin spectrum has a zero gap (the exception is the point \(\mathbf{\Gamma} = (0,0)\), where the spectrum is always gapless). Figure~\ref{fig_Cq_Diag} shows an example of a structure factor maximum, in this case on the diagonal of the Brillouin zone, corresponding to a helicoid \((q,q)\).

It turns out that, as in the classical limit, at \(S = 1/2\), in addition to the AFM, FM, and stripe spin structures, three types of helices are realized. They correspond to the position of the maximum of \(c_{\mathbf{q}}\) on one of the sides or the diagonal of the Brillouin zone (see Fig.~\ref{fig_PD_J123_quant}).

\begin{figure}[h]
  \centering
  \includegraphics[width=0.75\textwidth]{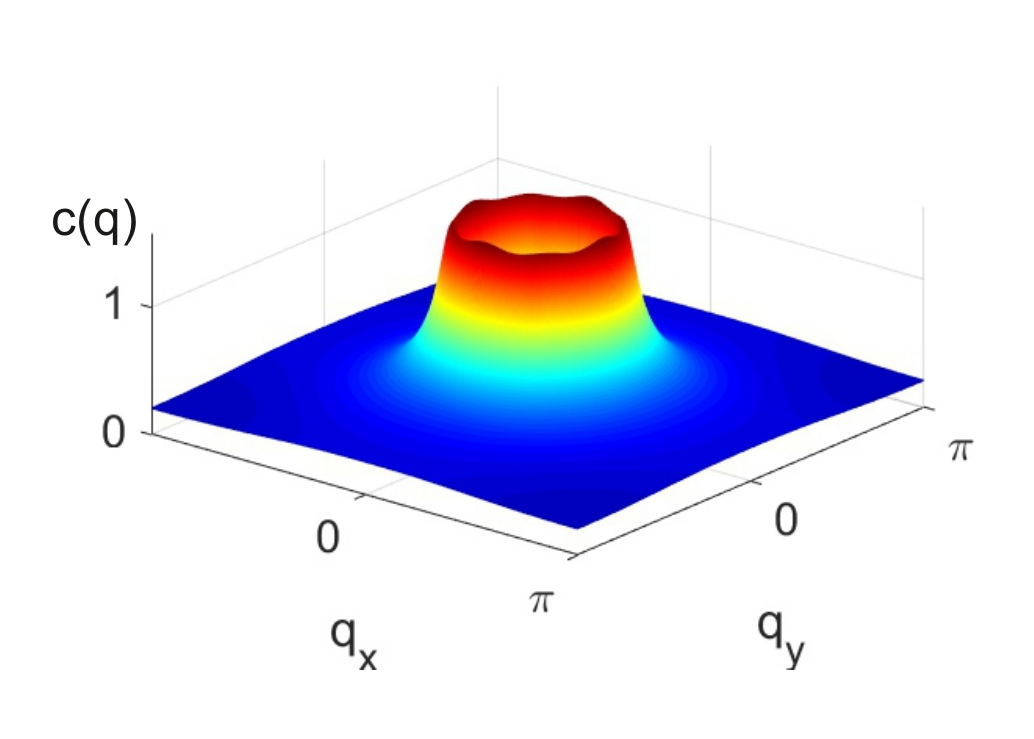}
  \caption{\small (Color online) Structure factor with almost perfect circular symmetry. \(T=0.02\), \(\varphi=160^{\circ}\), \(\psi=10^{\circ}\). The control point \((q,q)\) is close to that determining Fig.~\ref{fig_Cq_Diag}, but shifted closer to the FM point \(\mathbf{\Gamma} = (0,0)\). With this shift, the spreading of the sharp peak in Fig.~\ref{fig_Cq_Diag} leads to an almost circular structure. Unlike Fig.~\ref{fig_Cq_Diag}, the full Brillouin zone is shown here for clarity (data from \protect\cite{Valiul19_JPCM}).}
  \label{fig_Cq_Round}
\end{figure}

In the \(J_1\)-\(J_2\)-\(J_3\) model, another nontrivial feature appears that is not realized in the \(J_1\)-\(J_2\) and \(J_1\) models. Since the transition from one spin order motif to another occurs continuously, in the transition region, due to their interference, the most bizarre structures can arise, in particular, those possessing nearly circular symmetry. An example of such an isotropic helicoid is shown in Fig.~\ref{fig_Cq_Round}. The structure factor acquires a nearly isotropic form as the sharp maximum in Fig.~\ref{fig_Cq_Diag} spreads out, when the control point \((q,q)\) moves along the diagonal toward the FM point \(\mathbf{\Gamma} = (0.0)\). A similar effect is observed as the control point approaches the AFM point \(\mathbf{Q} = (\pi,\pi)\) along the diagonal. Such spin structures can be viewed as quantum helices `propagating in all directions.'

In the \(J_1\)-\(J_2\) model, a truncated analog of this effect is the gradual flow of the structure factor from one symmetric point to another as one moves along the \(J_1\)-\(J_2\) circle (see previous section).

\textit{Susceptibility and heat capacity.} Let us briefly mention the behavior of the spin susceptibility and heat capacity in the \(J_1\)-\(J_2\)-\(J_3\) model.

The addition of a third exchange does not qualitatively change the behavior of \(C_V\) and \(\chi(\mathbf{q},\omega = 0)\) overall. The heat capacity exhibits a standard temperature curve — a single maximum and zero asymptotics as \(T \to 0\) (the exception is an additional low-temperature maximum in the region of the stripe \(\leftrightarrow\) FM transition).

The static susceptibility \(\chi(\mathbf{q},\omega = 0)\), calculated at a specific symmetric point \(\mathbf{q}_0\) of the Brillouin zone, exhibits a maximum in the region of exchange parameters corresponding to the relevant spin structure.

One can mention a not entirely trivial effect — the susceptibility \(\chi(\mathbf{q},\omega = 0)\) calculated on the diagonal of the Brillouin zone, i.e., at any point with coordinates \(\mathbf{q} = (q_0,q_0)\), besides a maximum in the region of this point, also has a smaller but comparable maximum in the region of the point \(\mathbf{q} = (\pi - q_0,\pi - q_0)\). The latter corresponds to a helicoid of `similar' structure.

\subsubsection{Coexistence of two long-range orders}

\begin{figure}[h]
  \centering
  \includegraphics[width=0.6\textwidth]{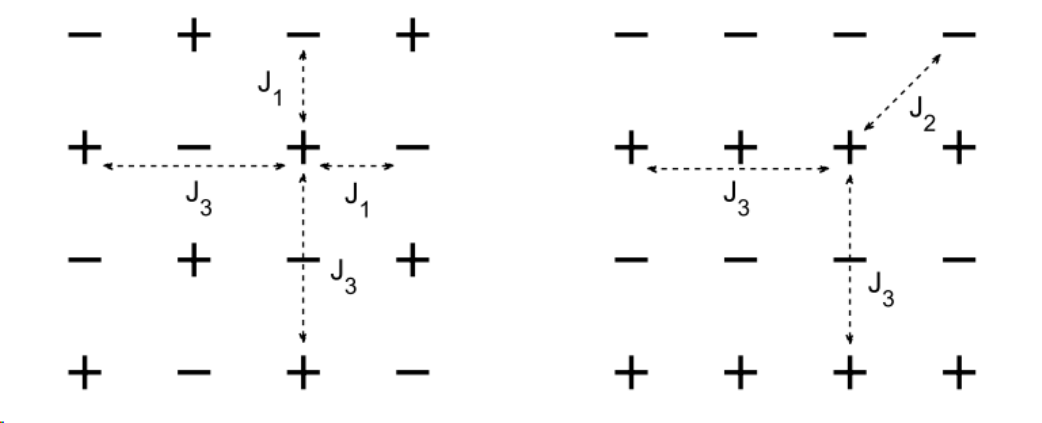}
  \caption{\small Negative, i.e., ferromagnetic, exchange \(J_3\) (at twice the lattice constant) favors both the antiferromagnetic and stripe phases. In the first case, AFM coupling is realized on the nearest neighbors; in the second, on the second nearest neighbors. In both cases, the coupling on the third neighbors is antiferromagnetic.}
  \label{fig_Byordered0}
\end{figure}

In the two-dimensional quantum \(J_1\)-\(J_2\)-\(J_3\) model, another nontrivial effect is possible: the coexistence of two different long-range orders at \(T = 0\) \cite{Mikhey20_PB}. Simple geometric considerations already indicate this possibility. As can be seen from Fig.~\ref{fig_Byordered0}, on a square lattice, a negative third exchange \(J_3\) favors both antiferromagnetic and stripe orders simultaneously.

\begin{figure}[h]
  \centering
  \includegraphics[width=0.7\textwidth]{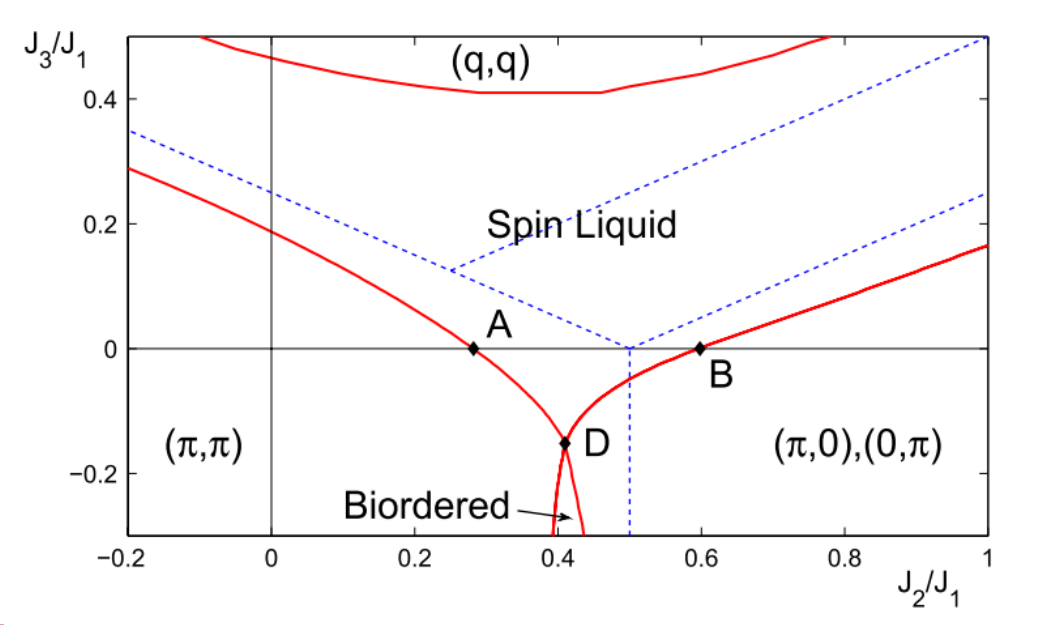}
  \caption{\small (Color online) Section of the phase diagram of the \(J_1\)-\(J_2\)-\(J_3\) model at \(T = 0\); an exchange parameterization different from form (\ref{Param_123}) is used. Red lines are boundaries of quantum phases: at the top — diagonal helicoid \(\mathbf{q} = (q,q)\), in the center — spin liquid, bottom left — AFM, bottom right — stripe long-range order, bottom center — phase with two orders. Dotted lines correspond to boundaries of classical phases: bottom — AFM and stripe, above — two helicoids. Data from \protect\cite{Mikhey20_PB}.}
  \label{fig_Byordered1}
\end{figure}

In the RGM, in the quantum case \(S = 1/2\), this leads to the appearance at \(J_1,J_2 > 0\) and a sufficiently large magnitude of \(J_3 < 0\) in a certain range of the parameter \(J_2/J_1\) of a state with two coexisting long-range orders — antiferromagnetic and stripe (see Fig.~\ref{fig_Byordered1}).

The structure factor \(c_{\mathbf{q}}\) of such a state has \(\delta\)-shaped peaks both at the AFM point of the Brillouin zone \(\mathbf{Q} = (\pi,\pi)\) and at the stripe points \(\mathbf{X} = (0,\pi), (\pi,0)\). Spin condensates of both symmetries are present (for details, see Section \ref{J1-J2}). The intensity of the AFM condensate decreases, while that of the stripe condensate increases, with increasing \(J_2/J_1\) from one side of the region with two condensates to the other. The spin excitation spectrum has zero gaps at all four high-symmetry points: \(\mathbf{\Gamma} = (0,0)\), \(\mathbf{X} = (0,\pi), (\pi,0)\), and \(\mathbf{Q} = (\pi,\pi)\).

Thus, the transition from the AFM to the stripe phase for positive and small negative \(J_3\) occurs through the spin-liquid phase, while for large negative \(J_3\) it occurs through a phase with two long-range orders.

In the classical limit, the described effect is impossible. This is intuitively obvious, but a rigorous proof, i.e., substituting ansatz (\ref{S_r}) into the classical expression for the energy followed by minimization, also yields the same answer.

Earlier in Sections \ref{J1-J2} and \ref{frust_helix}, the coexistence of different spin orders in the RGM was already mentioned. In particular, at \(T = 0\), the stripe phase is a quantum superposition of two long-range orders. The case described here differs in that two long-range orders of different symmetries coexist.

\subsubsection{\(J_1\)-\(J_2\)-\(J_3\) and experiment}

The above may seem like beautiful, but merely model speculation, since it is difficult to expect the exchange relation \(J_3 \sim J_1,J_2\) to be realized in experiment. This could perhaps be achieved in a suitable quantum simulator (textbook reviews \cite{Lewens07_AP,Bloch08_RMP,George14_RoMP} and a recent forecast review \cite{Fraxan23_Book_Chap}).

It is important that in some areas of the phase diagram, helicoids arise even at small \(J_3\). This makes the experimental realization of a \(J_1\)-\(J_2\)-\(J_3\) helix realistic. One possible example is (CuBr)Sr\(_2\)Nb\(_3\)O\(_{10}\). This is a quasi-two-dimensional compound with a square lattice in the plane. A neutron experiment \cite{Tsujim07_JPSJ,Tsujim08_PRB,Yusuf11_PRB,Ritter13_PRB} reveals a peak in the structure factor at the position \((3\pi/4, 0)\), which corresponds to a commensurate spin helix. There is no breaking of inversion symmetry, which makes the DMI scenario impossible.

The standard implementation of the RGM for the \(J_1\)-\(J_2\)-\(J_3\) model allows one to obtain the peak position at \((3\pi/4, 0)\) at low temperatures for a sufficiently small \(J_3\) (\(J_1 = -0.81\), \(J_2 = 0.56\), \(J_3 = 0.17\)) \cite{Mikhey20_PB}. In this case, the experimentally observed heat capacity and spin susceptibility of the system are reproduced with good accuracy.

\section[Relationship to other spherically symmetric approaches]{Relationship to other\\ spherically symmetric approaches}
\label{Relat}

In addition to the above-described approach, there exist other spherically symmetric methods that differ from the RGM. However, they are based on the same idea — description of a state with zero mean spin at a site by extracting spin-spin correlators from the Green's function. All these methods employ double-time retarded Green's functions, for which several canonical analysis algorithms are available. Essentially, each method represents an adaptation of a canonical algorithm to the constraints of spherical symmetry and zero on-site spin. All approaches, despite some mathematical differences, are conceptually close to one another, and it is often difficult to draw a strict boundary between them. Their results agree qualitatively, and often with high accuracy quantitatively. The essential dependences are essentially the same; to simplify, one can say that the difference in numerical answers reduces essentially to manipulating the vertex corrections.\footnote{Since this section is devoted not to specific results but to methods — analogues of the RGM — it will mention works related not only to the square lattice but also to other geometries.}

The approach described in the previous sections represents an adaptation of one of the standard algorithms — successive differentiation with respect to the `second' time and chain closure with the extraction of spin-spin correlators. Some approaches related to the RGM are briefly described below.

\subsection{Matrix projection method}
\label{Relat_Proj}

In principle, the Zwanzig-Mori-Tserkovnikov projection method \cite{Zwanzi60_JCP,Zwanzi61_PR,Zwanzi65_ARPC,Zwanzi72_PRA,Mori65_PTP,Mori65_PTPa,Tserko71_TMP_R,Tserko81_TMP_R} is universal, in the sense that its corresponding version can be used for states with both zero and nonzero average spin at the site. A possible adaptation of the scheme to the spherically symmetric case is presented below \cite{Baraba15_JL_R}. This variant is closest to the RGM, and with some simplifications, it directly transforms into the RGM. Therefore, it is described below in somewhat greater detail.

First, we present the basic equations of the method in general form.

Let us calculate the double-time retarded Green's functions for a set of operators \(\hat{A}_{i}\), (\(i= 1 \div n\))
\begin{equation}
G_{ij} = \langle \hat{A}_{i}|\hat{A}_{j}\rangle _{\omega +i\delta} =
-i\int\limits_{0}^{\infty }dt\,e^{i\omega t}
\langle \lbrack \hat{A}_{i}(t),\hat{A}_{j}^{+}]\rangle   \label{GreenF_Proj}
\end{equation}
Here, \(i\) is not the site index, as above, but the operator number. The operators \(\hat{A}_{i}\) form a vector \(\mathbf{A}=\{{\hat{A}_{1}\ldots \hat{A}_{n}}\}\).

Let us introduce matrix notation. We denote the \(n \times n\) matrix of Green's functions \(G_{ij}\) by \(\mathbf{G}\), the overlap matrix \(K_{ij}=\langle \lbrack \hat{A}_{i},\hat{A}_{j}^{+}]\rangle\) by \(\mathbf{K}=\langle \lbrack \mathbf{A},\mathbf{A}^{+}]\rangle\), and the frequency matrix by \(\mathbf{\omega} = \omega \mathbf{I}\), \((\mathbf{I)}_{ij} = \delta _{ij}\).

We now introduce the vector \(\mathbf{B} = {\hat{B}_{1}\ldots \hat{B}_{n}}\) of operators \(\hat{B}_{i}\) arising in the chain of equations for the Green's functions \(\mathbf{B} = [\mathbf{A},\hat{H}]\) (\(\hat{H}\) is the Hamiltonian), as well as the energy matrix \(\mathbf{\Omega} = \langle [\mathbf{B},\mathbf{A}^{+}]\rangle\), determined by the commutator means \(\Omega_{ij} = \langle \lbrack \hat{B}_{i},\hat{A}_{j}^{+}]\rangle\).

Then it can be shown that the analog of the Dyson equation for \(\mathbf{G}\) has the following form
\begin{equation}
\mathbf{G\times }\left( \mathbf{\omega -K}^{-1}\mathbf{\Omega -K}^{-1}
\mathbf{L}^{irr}\right) = \mathbf{K},  \label{Dyson}
\end{equation}
where the matrix of irreducible parts is
\begin{equation}
\mathbf{L}^{irr} = \langle \mathbf{B|B}^{+}\rangle -\langle \mathbf{B|A}^{+}
\rangle \mathbf{G}^{-1}\langle \mathbf{A|B}^{+}\rangle ,
\end{equation}
and its elements are
\begin{equation}
L_{ij}^{irr} = \langle {\hat{B}_{i}}|{\hat{B}_{j}^{+}}\rangle -
\langle {\hat{B}_{i}}|\hat{A}_{s}^{+}\rangle (\hat{G}^{-1})_{st}\langle \hat{A}_{t}|
{\hat{B}_{j}^{+}}\rangle .
\end{equation}

We now return to the spherically symmetric approach. We introduce notations for successive commutators
\begin{equation}
D_{\mathbf{n}}^{z} = [S_{\mathbf{n}}^{z},\hat{H}];\quad
W_{\mathbf{n}}^{z} = [[S_{\mathbf{n}}^{z},\hat{H}],\hat{H}] =
[D_{\mathbf{n}}^{z},\hat{H}]
\end{equation}

Then the first two steps in the chain of equations for the Green's function (\ref{First step}) and (\ref{Second step}) can be written in compact form
\begin{equation}
\omega \langle S_{\mathbf{n}}^{z}|S_{\mathbf{m}}^{z}\rangle =\langle D_{%
\mathbf{n}}^{z}|S_{\mathbf{m}}^{z}\rangle
\label{First step_comp}
\end{equation}
\begin{equation}
\omega \langle D_{\mathbf{n}}^{z}|S_{\mathbf{m}}^{z}\rangle =\langle \lbrack
D_{\mathbf{n}}^{z},S_{\mathbf{m}}^{z}]\rangle +\langle W_{\mathbf{n}}^{z}|S_{%
\mathbf{m}}^{z}\rangle
\label{Second step_comp}
\end{equation}
Here, frequency indices are omitted for simplicity.

Next, in the standard RGM algorithm (described above in Section \ref{Algor}), spin-spin correlation functions with vertex corrections are extracted from the three-site terms on the right-hand side of (\ref{Second step_comp}). The projection method implements a slightly different approach. After the Fourier transform, the system of equations (\ref{First step_comp}), (\ref{Second step_comp}) takes the form
\begin{equation}
\omega \langle S_{\mathbf{q}}^{z}|(S_{\mathbf{q}}^{z})^{+}\rangle =
\langle D_{\mathbf{q}}^{z}|(S_{\mathbf{q}}^{z})^{+}\rangle
\label{First step_bis_Fur}
\end{equation}
\begin{equation}
\omega \langle D_{\mathbf{q}}^{z}|(S_{\mathbf{q}}^{z})^{+}\rangle =
F_{\mathbf{q}}+\langle W_{\mathbf{q}}^{z}|(S_{\mathbf{q}}^{z})^{+}\rangle
\label{Second step_bis_Fur}
\end{equation}

The specific expressions for \(D_{\mathbf{q}}^{z}\), \(F_{\mathbf{q}}\) and \(W_{\mathbf{q}}^{z}\) depend on the model (\(J_1\), \(J_1\)-\(J_2\), or \(J_1\)-\(J_2\)-\(J_3\)), and we will not present them here. In particular, \(F_{\mathbf{q}}\) is the numerator of the Green's function (\ref{Gz}). We emphasize that equations (\ref{First step_bis_Fur}), (\ref{Second step_bis_Fur}) are exact at this stage.

Now let us move on to implementing the projection method. We will limit ourselves to a basis of two operators; then, their choice is obvious: \(\hat{A}_{1}=S_{\mathbf{q}}^{z}\), \(\hat{A}_{2}=D_{\mathbf{q}}^{z}\). In this basis, the quantities in (\ref{Dyson}) take the form
\begin{equation}
\mathbf{K}=
\begin{vmatrix}
0 & F_{\mathbf{q}} \\
F_{\mathbf{q}} & 0%
\end{vmatrix}%
\label{Proj_K}
\end{equation}
\begin{equation}
\mathbf{K}^{-1}\mathbf{\Omega }=
\begin{vmatrix}
0 & \omega _{\mathbf{q}}^{2} \\
1 & 0
\end{vmatrix}
\end{equation}
where \(\omega _{\mathbf{q}}^{2}=F_{\mathbf{q}}^{-1}\langle W_{\mathbf{q}}^{z}| (D_{\mathbf{q}}^{z})^{+}\rangle\), and the matrix of irreducible parts is
\begin{equation}
\mathbf{L}^{irr} =
\begin{vmatrix}
0 & 0 \\
0 & \langle W_{\mathbf{q}}^{z}|(W_{\mathbf{q}}^{z})^{+}\rangle ^{irr}
\end{vmatrix}
\end{equation}
Then on the left-hand side of (\ref{Dyson})
\begin{equation}
\left( \mathbf{\omega -K}^{-1}\mathbf{\Omega -K}^{-1}\mathbf{L}^{irr}\right) =
\begin{vmatrix}
\omega  & -\omega _{\mathbf{q}}^{2}-F_{\mathbf{q}}^{-1}\langle W_{\mathbf{q}
}^{z}|(W_{\mathbf{q}}^{z})^{+}\rangle ^{irr} \\
-1 & \omega
\end{vmatrix}
\end{equation}
And finally, equation (\ref{Dyson}) takes the form
\begin{equation}
\begin{vmatrix}
G_{11} & G_{12} \\
G_{21} & G_{22}
\end{vmatrix}
\begin{vmatrix}
\omega  & -\omega _{\mathbf{q}}^{2}-F_{\mathbf{q}}^{-1}
\langle W_{\mathbf{q}}^{z}|(W_{\mathbf{q}}^{z})^{+}\rangle ^{irr} \\
-1 & \omega
\end{vmatrix} =
\begin{vmatrix}
0 & F_{\mathbf{q}} \\
F_{\mathbf{q}} & 0
\end{vmatrix}
\end{equation}
The solution of the pair of equations related to the top row of the matrix \(\mathbf{G}\)
\begin{eqnarray}
\omega G_{11}-G_{12} &=&0 \\
-G_{11}\left[ (\omega _{\mathbf{q}}^{2}+F_{\mathbf{q}}^{-1}\langle
W_{\mathbf{q}}^{z}|(W_{\mathbf{q}}^{z})^{+}\rangle ^{irr})\right]
&=&F_{\mathbf{q}}
\end{eqnarray}
yields the Green's function \(G_{11} = G^{zz}(\omega ,\mathbf{q})\)\footnote{Here, for clarity, both superscripts are retained for the Green's function \(G^{zz}.\)}
\begin{equation}
G_{11}= G^{zz}(\omega ,\mathbf{q}) =
\frac{F_{\mathbf{q}}}{\omega ^{2}-\left[ (\omega _{\mathbf{q}}^{2} +
F_{\mathbf{q}}^{-1}\langle W_{\mathbf{q}}^{z}|(W_{\mathbf{q}}^{z})^{+}\rangle
^{irr})\right] }
\label{Proj_GF}
\end{equation}

This is the Green's function of interest.

Here, two remarks are necessary. First, when deriving relations (\ref{Proj_K})–(\ref{Proj_GF}), the constraints of the RGM were strictly observed, i.e., the requirements of spherical symmetry and zero average spin at the site. Second, all these relations remain exact.

In expression (\ref{Proj_GF}), the quantity \(\omega _{\mathbf{q}}\) has the meaning of a mean-field spectrum of spin excitations, and accounting for \(\langle W_{\mathbf{q}}^{z}|(W_{\mathbf{q}}^{z})^{+}\rangle^{irr})\) leads to going beyond the mean field, in particular, to the appearance of damping.

It can be shown \cite{Baraba15_JL_R} that if a series of simplifications is made in the sequential calculation of \(\omega _{\mathbf{q}}\), the resulting expression reduces to the standard RGM spectrum (with vertex corrections). When attempting to go further, a \(\mathbf{q}\)-dependence of the vertex corrections inevitably arises. However, directly determining its form encounters significant mathematical difficulties.

These difficulties can be partially overcome in a semi-phenomenological approach by assuming a physically natural \(\mathbf{q}\)-dependence of the vertices (details in \cite{Baraba15_JL_R}) and including it in the full self-consistency procedure. Even this simplest variant significantly refines the quantitative results of the RGM.

This concerns, in particular, the zero-temperature limit in the model with a single exchange, i.e., the point \(J_2 = J_3 = 0\), \(T = 0\), which has been studied in detail by alternative, mainly numerical, methods. Here, the basic RGM variant, depending on the choice of vertices, gives good agreement with numerical data either only for the energy (i.e., for the first correlator), or only for the effective magnetization (for the correlator at infinity). Including the \(\mathbf{q}\)-dependence of the vertices eliminates this problem.

Some other ways to `fine-tune' the RGM are described in Section~\ref{Renorm}.

\subsection{Continued fraction method and others}

The continued fraction method is based on the use of Mori projection operator techniques \cite{Mori65_PTP,Mori65_PTPa} for double-time retarded Green's functions of the form (\ref{GreenF}) (similar to Sections \ref{Algor} and \ref{Relat_Proj}). This method was applied within the \(t-J\) model on a triangular \cite{Rubin08_PLA} and square \cite{Sherma02_PRB,Sherma05_PLA} lattice, as well as for various variations of the Heisenberg model on a triangular lattice (\(S=1/2\) \(J_1\) \cite{Rubin05_PLA}, \(S=1\) \(J_{1}\)–\(J_{2}\) \cite{Rubin10_PLA}, and \(S=1\) \(J_{1}\)–\(J_{3}\) \cite{Rubin12_PLAa}).

Within this method, the spin Green's function can be represented as a continued fraction (see, e.g., \cite{Sherma02_PRB}):
\begin{equation}\label{cep_drop}
	G^{zz}\left( \omega ,\mathbf{q}\right) =\frac{F_{\mathbf{q}}}{\omega -E_{0}-%
		\frac{V_{0}}{\omega -E_{1}-\frac{V_{1}}{\ddots }}},
\end{equation}
where in the general case the elements \(E_i\) and \(V_i\) are calculated using a recursive procedure:
\begin{equation}
	[A_n, H] = E_n A_n + A_{n+1} + V_{n-1} A_{n-1},
\end{equation}
\begin{equation}
	E_n = \left( [A_n, H], A^\dagger_n \right) \left( A_n, A^\dagger_n \right)^{-1},
\end{equation}
\begin{equation}
	V_{n-1} = \left(A_n, A^\dagger_n \right) \left( A_{n-1}, A^\dagger_{n-1} \right)^{-1},
\end{equation}
The operators \(A_i\) form an orthogonal set satisfying \((A_i, A^\dagger_j) \propto~\delta_{ij}\), where the notation is introduced:
\begin{equation}
	(A, B) = i \int_0^\infty dt \, \left\langle [A(t), B] \right\rangle.
\end{equation}

For spin models, the starting point of the recursion is chosen as
\begin{equation}
	V_{-1} = 0, \quad A_0 =S_\textbf{q}^z, \quad n = 0, 1, 2, \dots.
\end{equation}

Further calculations in the cited works reduce to truncating the continued fraction at the second step, i.e., choosing the approximation \(V_1 = 0\). Such an approximation is essentially equivalent to neglecting four-site (and higher) Green's functions. The resulting three-site Green's functions of the form \(S^z_\textbf{l}S^+_\textbf{m}S^-_\textbf{n}\) are decoupled in the usual RGM manner — by introducing vertex corrections. The method strictly satisfies the RGM conditions due to taking into account the requirement of zero magnetization at the site:
\begin{equation}\label{cep_drop2}
	\left\langle S^z_\textbf{n} \right\rangle = \frac{1}{2} - \left\langle S^{-}_\textbf{n} S^{+}_\textbf{n} \right\rangle = 0.
\end{equation}
As can be easily shown, this restriction is equivalent to the constraint condition (in particular, for \(S = 1/2\), \(\left\langle S^2_\textbf{n} \right\rangle = \frac{3}{4}\)).

The system of equations for the spin correlators together with the constraint condition (\ref{cep_drop2}) is solved self-consistently. If only one vertex correction is chosen, the system of equations is closed and its solution is possible without additional conditions. However, if two vertex corrections are used, additional freedom appears: the number of variables exceeds the number of equations. In this case, it becomes necessary to introduce an additional constraint for the solution, for example, the requirement to reproduce the results of exact diagonalization calculations for the value of the correlator \(c_g\) \cite{Rubin05_PLA}.

In the literature, one can also find references to the method of differentiating the Green's function with respect to both times and the method of formally exact solution (general reviews can be found in \cite{Plakid11_TMF_R,Rudoy11_TMF_R}), modified to account for spherical symmetry. These approaches (not described in detail here) ultimately reduce to similar decouplings and the same expression for the Green's function as everywhere above.

\section{`Fine-tuning' of calculations}
\label{Renorm}

As already mentioned in Section \ref{pro_contra}, the RGM method, although quite sophisticated and cumbersome, allowing the description of many effects in low-dimensional magnets, is still a mean-field approach. It does not allow the determination of the damping of spin excitations within its basic version.

This drawback can only be corrected phenomenologically \cite{Mikhey06_PLA,Baraba07_PLA,Mikhee07_JETP_R,Baraba11_TMP_R} (or, more correctly, semi-phenomenologically, see explanations below). One such approach was mentioned above in Section \ref{Relat_Proj}. Another possible path is described below.

A real and imaginary renormalization is introduced into the expression for the Green's function (\ref{Gz}), and it takes the form
\begin{equation}
G^{z}(\mathbf{q},\omega ) = \frac{F_{\mathbf{q}}}
{\omega ^{2}-\omega _{\mathbf{q}}^{2}-M(\mathbf{q},\omega )}, \quad
M(\mathbf{q},\omega ) = \mathrm{Re} M+i\ \mathrm{Im} M.
\label{Gz_renorm_1}
\end{equation}

The form (\ref{Gz_renorm_1}) seems intuitively obvious. However, it is a formally exact expression for the Green's function \(\langle S_{\mathbf{q}}^{z}|S_{-\mathbf{q}}^{z}\rangle _{\omega }\). It can be obtained by the method of irreducible Green's functions (details can be found in Section \ref{Relat}). Here \(M(\mathbf{q},\omega)\) is the Fourier transform of a new complex Green's function whose analytical properties are the same as those of \(G^{z}(\mathbf{q},\omega )\) \cite{Baraba95_PLA}. The function \(M(\mathbf{q},\omega)\) corresponds to a three-site irreducible retarded Green's function and describes the decay of a spin excitation.

Direct calculation of \(M(\mathbf{q},\omega)\) even for the simplest models encounters significant mathematical difficulties, so one has to resort to a semi-phenomenological consideration.

Let us write (\ref{Gz_renorm_1}) in a slightly different form
\begin{equation}
G^{z}(\mathbf{q},\omega ) = \frac{F_{\mathbf{q}}}
{\omega ^{2}-\widetilde{\omega }_{\mathbf{q}}^{2}+i\omega \gamma (\mathbf{q},\omega )}
\label{Gz_renorm_2}
\end{equation}

In (\ref{Gz_renorm_2}), \(\widetilde{\omega }_{\mathbf{q}}^{2} = \omega _{\mathbf{q}}^{2}-\mathrm{Re}M_{\mathbf{q}}\) gives an effective renormalized spectrum of spin waves (if one neglects the dependence of \(\mathrm{Re}M_{\mathbf{q}}\) on \(\omega\)). The imaginary part of the renormalization is an odd function of \(\omega\) and is written in (\ref{Gz_renorm_2}) as \(\mathrm{Im} M = -\omega \gamma (\mathbf{q},\omega )\), which is convenient for specific calculations (thus, \(\gamma (\mathbf{q},\omega)\) is an even function of \(\omega\)).

\textit{A necessary clarification.} The term `semi-phenomenological' is used in the following sense. On the one hand, the functional form of the real and imaginary renormalizations and the coefficients in them are not calculated directly but are introduced `from outside', and are determined taking into account additional considerations and constraints. But, on the other hand, the contributions from the renormalizations are included on a general basis in the full self-consistency procedure.

Introducing renormalizations makes it possible to improve the quantitative results of the RGM, and sometimes to describe effects that are in principle inaccessible with the original Green's function (\ref{Gz}). Omitting technical details and limiting ourselves to basic references, we give two examples, both related to the \(J_1\)-\(J_2\) model.

\textit{Width of the spin-liquid region.} At \(T = 0\), the basic RGM variant correctly determines the sequence of phases with and without long-range order on the entire \(J_1\)-\(J_2\) circle (see Fig.~\ref{fig_Circle_Quant}).

This is also true for the first quadrant of the circle \(J_1 > 0\), \(J_2 > 0\). For \(J_1 \gg J_2\), the RGM predicts an AFM phase with long-range order; for \(J_1 \sim J_2\), a spin liquid (SL) without long-range order; and for \(J_1 \ll J_2\), a stripe phase with long-range order.

The AFM-SL and SL-stripe transitions on a square lattice are canonical, textbook examples of quantum phase transitions \cite{Sachde11_Book}, and have therefore been studied in detail by various analytical and numerical methods. The transition points, and consequently the width of the spin-liquid region, are considered reliably established. The RGM with the Green's function (\ref{Gz}) overestimates the width of this region.

However, even the simplest renormalization of the form (\ref{Gz_renorm_2}) with \(\mathrm{Re}M=0\), \(\gamma(\mathbf{q},\omega ) = \mathrm{Const}\) at moderate values of \(\gamma\) returns the SL region to consensus values.

\textit{Scaling of the spin susceptibility.} Even at the early stage of studying HTSC in quasi-two-dimensional cuprates, scaling of the susceptibility was discovered (\cite{Kastne98_RMP}, see also the corresponding section in \cite{Plakida10_Book_Part}).

The scaling law refers to the so-called `local spin susceptibility' \(\chi _{2D}(\omega ,T)\) — the imaginary part of the susceptibility integrated over quasimomentum
\begin{equation}
\chi _{2D}(\omega ,T)=\int d\mathbf{q\,}\mathrm{Im}
\chi (\mathbf{q},\omega,T).
\end{equation}

It turns out that the dependence of \(\chi _{2D}\) on \(\omega\) and \(T\) over a wide range of dopings is well described by the following expression
\begin{equation}
\frac{\chi _{2D}(\omega ,T)}{\chi _{2D}(\omega ,T\rightarrow 0)} =
f(\frac{\omega }{T}),
\end{equation}
where the scaling function \(f\) is usually approximated by the arctangent of a cubic polynomial with coefficients depending on doping.

It is easy to show that in the basic RGM variant, without damping of spin excitations, this effect cannot be described in principle. But already a simple renormalization of the form (\ref{Gz_renorm_2}), where \(\mathrm{Re}M=0\) and the damping parameter depends only on temperature (and linearly), makes it possible to reproduce the scaling. Moreover, in this case, the problem can even be solved analytically for fixed values of the exchanges. As for the real renormalization, it allows a more quantitatively accurate description of the experiment for \(\chi _{2D}\).

Note that in more complex cases, the linear dependence of \(\gamma\) on temperature, necessarily following from scaling, serves as an additional constraint when choosing renormalizations.

\section{RGM and quantum entanglement}
\label{Kug_Khom}

The RGM can be applied not only to the Heisenberg model in its various variants but also to more complex structures with localized spins. This includes, in particular, the spin-orbital model (spin-pseudospin, the Kugel-Khomskii model \cite{Kugel73_JETP_R,Kugel82_UFN_R}).

Research on this model, proposed back in the 1970s, has recently experienced a surge in activity. This is due to two circumstances.

First, the orbital waves predicted by the model were finally experimentally detected in this century \cite{Saitoh01_N,Gruenin02_N,Saitoh02_N,Ishiha05_NJP}.

Second, the spin-orbital model turned out to be an extremely convenient testing ground for studying many-particle quantum entanglement \cite{Chen07_PRB,You12_PRB,Brzezi14_PRL,You15_PRB,Gotfry20_PRR,Mohapa22_JPCM,Valiul20_PRB,Valiul23_SPC}. In this area, analytical progress is almost impossible. Obtaining a numerical result is also difficult, since it is necessary to determine the density matrix of a macroscopic system. The features of the spin-pseudospin model — a natural division into two subsystems, the mathematical identity of the spin and pseudospin operators, and the availability of high-level spin computational packages — simplify this work. But even here, resource requirements force one to limit oneself to a small system size.

Analysis within the RGM, although it does not allow one to strictly determine the degree of entanglement, makes it possible to identify `suspicious regions' and qualitatively predict the expected results \cite{Kagan14_JL_R,Valiul19_JL_R}. And it is all the more suitable for describing the standard magnetic and thermodynamic properties of the model.

The Hamiltonian of the spin-pseudospin model has the form \cite{Kugel73_JETP_R,Kugel82_UFN_R}
\begin{equation}
\hat{H}=J\sum_{\langle \mathbf{i},\mathbf{j}\rangle } \hat{\mathbf{S}}_{\mathbf{i}}\hat{\mathbf{S}}_{\mathbf{j}} +
I\sum_{\langle \mathbf{i},\mathbf{j}\rangle } \hat{\mathbf{T}}_{\mathbf{i}}\hat{\mathbf{T}}_{\mathbf{j}} +
K\sum_{\langle \mathbf{i},\mathbf{j}\rangle }\left( \hat{\mathbf{S}}_{\mathbf{i}} \hat{\mathbf{S}}_{\mathbf{j}}\right) \left( \hat{\mathbf{T}}_{\mathbf{i}} \hat{\mathbf{T}}_{\mathbf{j}}\right)   \label{Hamilt_KK}
\end{equation}
At each lattice site, a spin \(S = 1/2\) and a pseudospin \(T = 1/2\) are fixed. The internal interactions in the spin and pseudospin subsystems are of Heisenberg type, the interaction between subsystems is the last term in (\ref{Hamilt_KK}). As in Hamiltonian (\ref{Hamilt}), \(\left\langle \mathbf{i,j}\right\rangle\) denotes summation over pairs of nearest neighbor sites.

Expression (\ref{Hamilt_KK}) represents the most symmetric, \(SU(2)\times SU(2)\) form of the model; in other variants, the pseudospin part may be different.

Let us consider low-dimensional (1D chain and 2D square) lattices at \(T > 0\). According to the Mermin-Wagner theorem \cite{Mermin66_PRL}, for uncoupled subsystems (\(K = 0\)), long-range order is impossible. It is natural to assume that with the inclusion of intersubsystem interaction \(K \neq 0\), the role of temperature and quantum fluctuations only increases, and there is no long-range order. This justifies the application of the RGM scheme.

Thus, the initial assumptions are conceptually the same as in the Heisenberg model: single-site averages for spin and pseudospin are zero \(\langle\hat{{S}}_{\mathbf{i}}\rangle = \langle \hat{{T}}_ {\mathbf{i}}\rangle =0\), and correlation functions \(\langle \hat{S}_{\mathbf{i}}^{\alpha}\hat{S}_{\mathbf{j}}^{\beta}\rangle\), \(\langle \hat{T}_{\mathbf{i}}^{\alpha}\hat{T}_{\mathbf{j}}^{\beta}\rangle\) and \(\langle \hat{S}_{\mathbf{i}}^{\alpha}\hat{T}_{\mathbf{j}}^{\beta}\rangle\) are zero for \(\alpha \neq \beta\) and independent of \(\alpha\) for \(\alpha = \beta\).

It is necessary to consider three interconnected Green's functions: spin-spin \(\langle S_{\mathbf{q}}^{z}|S_{-\mathbf{q}}^{z}\rangle_{\omega}\), spin-pseudospin \(\langle T_{\mathbf{q}}^{z}|S_{-\mathbf{q}}^{z}\rangle_{\omega}\), and pseudospin-pseudospin \(\langle T_{\mathbf{q}}^{z}|T_{-\mathbf{q}}^{z}\rangle_{\omega}\). It can be shown \cite{Pati98_PRL} that the effects of the interaction of spin and orbital degrees of freedom, including entanglement \cite{Lundgr12_PRB,You12_PRB}, are most clearly and vividly expressed at \(J = I > 0\) and \(K < 0\). For \(J = I\), obviously, \(\langle S_{\mathbf{q}}^{z}|S_{-\mathbf{q}}^{z}\rangle_{\omega} = \langle T_{\mathbf{q}}^{z}|T_{-\mathbf{q}}^{z}\rangle_{\omega}\), and two Green's functions remain.

However, even with these simplifications, calculations according to the scheme (\ref{First step})–(\ref{Second step}) cannot be carried out, since commutation of the spin (pseudospin) operator with the last term in (\ref{Hamilt_KK}) would go too far, and it would not be possible to close the chain of equations of motion. Nevertheless, by simplifying the intersubsystem interaction term in the spirit of the mean field (while preserving intrasite spin-pseudospin interactions, details see in \cite{Kagan14_JL_R}), one can obtain a closed system, the solution of which leads to nontrivial results.

\begin{figure}[h]
  \centering
  \textbf{a} \includegraphics[width=0.42\textwidth]{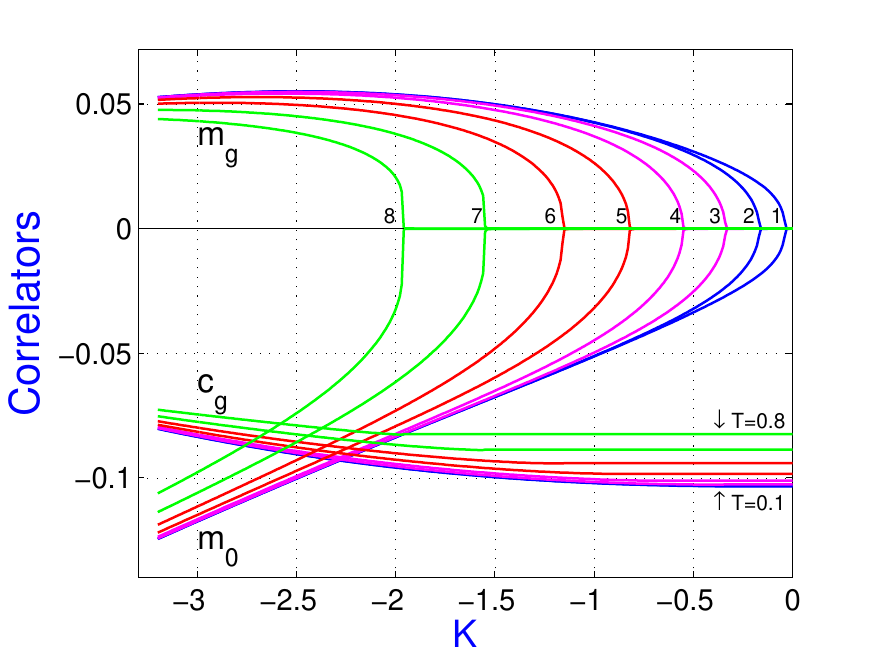}
  \textbf{b} \includegraphics[width=0.4\textwidth]{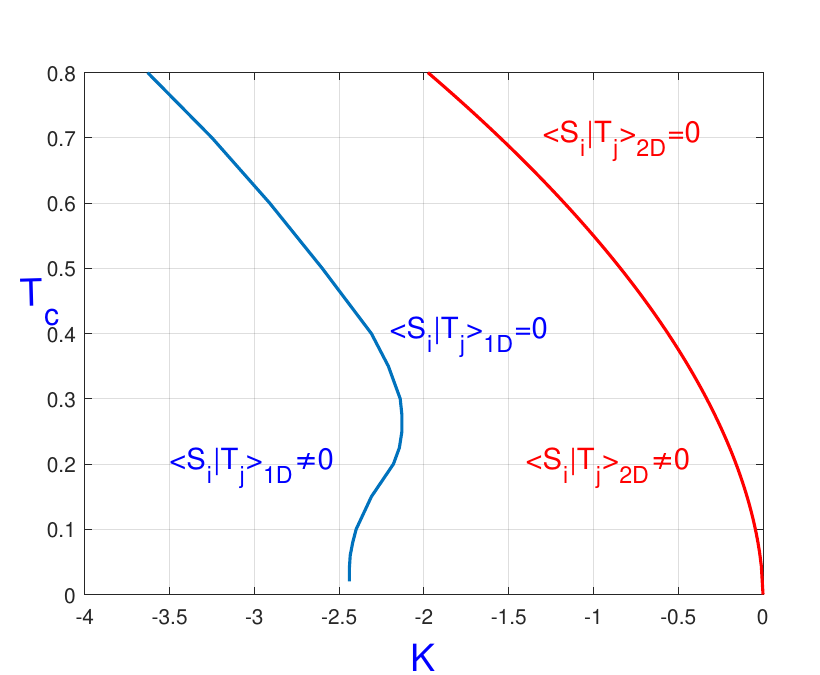}
  \caption{\small (Color online) \textbf{a.} 2D lattice. Dependence of the spin-spin correlator \(c_{g}\) on nearest neighbors, and spin-pseudospin correlators — intrasite \(m_{0}\) and intersite \(m_{g}\) — on temperature and intersubsystem exchange \(K\). The curves forming the `platypus nose' are \(m_{0}\) (\(m_{0}<0\)) and \(m_{g}\) (\(m_{g}>0\)). Numbers \(1\div8\) number the temperature. 1 — \(T=0.1\), 2 — \(T=0.2\), etc. The lower curves are \(c_{g}\) (boundary values of \(T\) are indicated). \textbf{b.} Phase diagram of the model for 1D and 2D. The boundaries of regions with zero and nonzero spin-pseudospin correlators \(m_{0}\) and \(m_{g}\) are shown (i.e., in the figure \(\mathbf{j = i,i+g}\)). The features of the one-dimensional case are noteworthy (see text) (data from \protect\cite{Kagan14_JL_R,Valiul19_JL_R}).}
  \label{fig_KK_1}
\end{figure}

\begin{figure}[h]
  \centering
  \textbf{a} \includegraphics[width=0.45\textwidth]{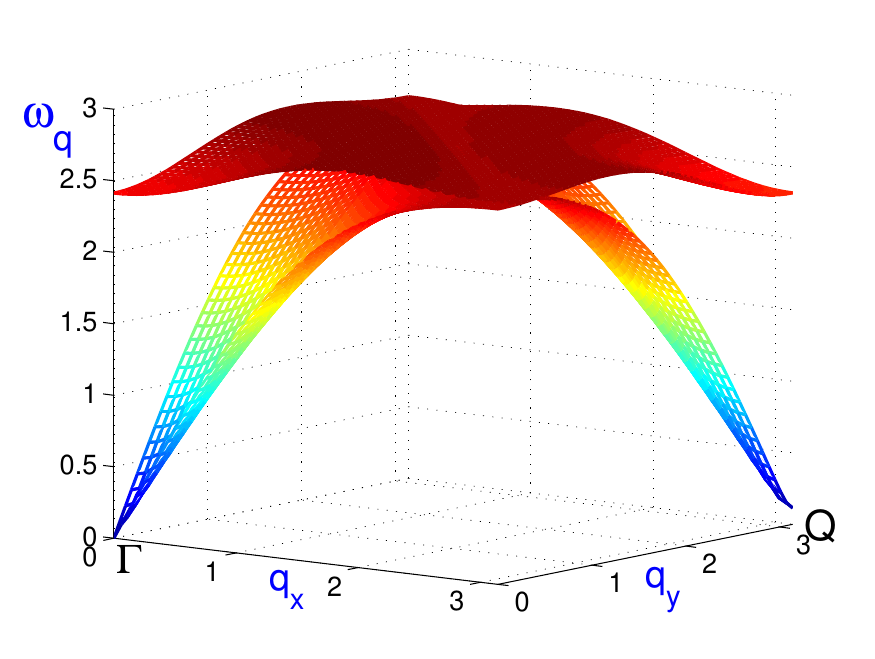}
  \textbf{b} \includegraphics[width=0.4\textwidth]{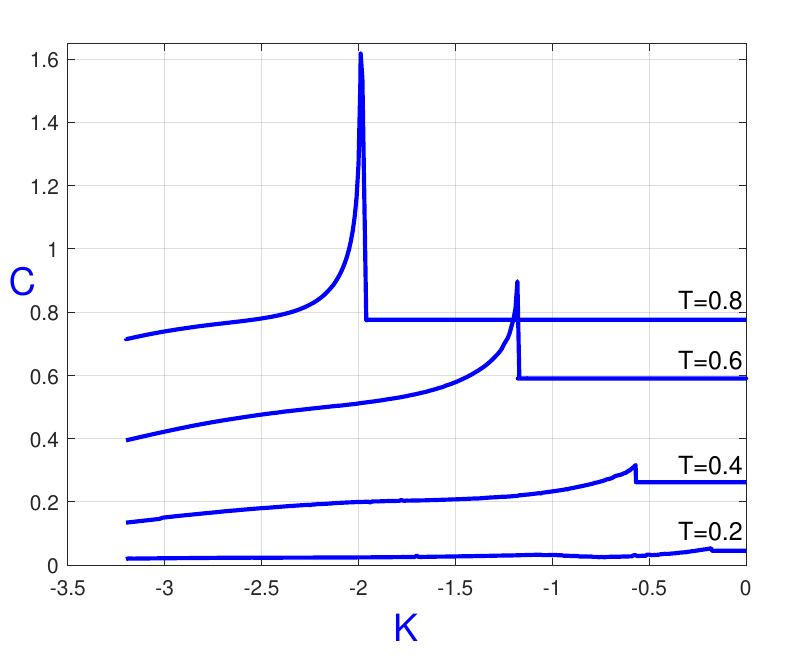}
  \caption{\small (Color online) \textbf{a.} 2D lattice. Spectra of elementary excitations \(\omega_{ac}(\mathbf{q})\) and \(\omega_{opt}(\mathbf{q})\) at \(T = 0.3\) and strong splitting \(K = -3.0\). The upper sections of the spectrum branches form an almost dispersionless region. A quarter of the full Brillouin zone is shown. \textbf{b.} 2D lattice. Dependence of the heat capacity at fixed temperature on the magnitude of the intersubsystem interaction \(K\). Horizontal sections correspond to zero values of the spin-pseudospin correlators (data from \protect\cite{Kagan14_JL_R,Valiul19_JL_R}).}
  \label{fig_KK_2}
\end{figure}

Taking into account the interaction of subsystems, the spin-spin \(G^{z}(\mathbf{q},\omega ) = \langle S_{\mathbf{q}}^{z}|S_{-\mathbf{q}}^{z}\rangle _{\omega}\) and spin-pseudospin Green's functions take the form
\begin{equation}
G^{z}(\mathbf{q},\omega ) = \langle S_{\mathbf{q}}^{z}|S_{-\mathbf{q}}^{z}\rangle _{\omega} = \frac{F_{ac}(\mathbf{q})}{\omega ^{2}-\omega_{ac}^{2}(\mathbf{q})}+
\frac{F_{opt}(\mathbf{q})}{\omega ^{2}-\omega_{opt}^{2}(\mathbf{q})},  \label{Gz_KK}
\end{equation}
\begin{equation}
R^{z}(\mathbf{q},\omega )=\langle T_{\mathbf{q}}^{z}|S_{-\mathbf{q}}^{z}\rangle _{\omega} = \frac{F_{ac}(\mathbf{q})}{\omega ^{2}-\omega_{ac}^{2}(\mathbf{q})}-
\frac{F_{opt}(\mathbf{q})}{\omega ^{2}-\omega_{opt}^{2}(\mathbf{q})},  \label{Rz_KK}
\end{equation}
Specific expressions for the numerators and spectra in (\ref{Gz_KK}) and (\ref{Rz_KK}), not essential for the further presentation, can be found in \cite{Kagan14_JL_R}.

The main results of the RGM for low-dimensional spin-pseudospin models are as follows. At a fixed temperature \(T > 0\), with increasing modulus of the intersubsystem interaction \(| K |\) (recall that \(K < 0\)), starting from a certain boundary value \(K_c\), nonzero spin-pseudospin correlators appear — intrasite \(m_{0}=\langle S_{\mathbf{i}}^{z}T_{\mathbf{i}}^{z}|\rangle\) and intersite \(m_{g}=\langle S_{\mathbf{i}}^{z}T_{\mathbf{ig}}^{z}|\rangle\), for 2D see Fig.~\ref{fig_KK_1}a.

The boundaries of regions with zero and nonzero spin-pseudospin correlators in 1D and 2D are shown in Fig.~\ref{fig_KK_1}b. Noteworthy is the counterintuitive behavior of the boundary in the one-dimensional case — in a certain range of the intersubsystem interaction, a reversible transition is observed with increasing temperature.

Simultaneously with the appearance of intersubsystem correlators, the excitation spectrum splits into acoustic and optical branches (see Fig.~\ref{fig_KK_2}a (2D)). For small \(| K |\), the splitting is noticeable mainly at the symmetric points of the Brillouin zone \(\mathbf{\Gamma}\) and \(\mathbf{Q}\). For large \(| K |\), the upper sections of the spectrum branches form an almost dispersionless region (see the same Fig.~\ref{fig_KK_2}a).

At the transition point, the heat capacity of the system also experiences a jump, the larger the greater the critical value \(| K |\) (see Fig.~\ref{fig_KK_2}b (2D)). Both susceptibilities, spin-spin \(\chi _{ss}(\mathbf{q},\omega )=-G^{z}(\mathbf{q},\omega )\) and spin-pseudospin \(\chi _{st}(\mathbf{q},\omega )=-R^{z}(\mathbf{q},\omega )\), also undergo jumps.

The set of these indicators can apparently be considered as witnesses of entanglement, although the theory of entanglement witnesses in condensed matter is still far from complete (see recent review \cite{Laurel24_AQT}).

In the exact numerical solution of the problem for a small one-dimensional chain \cite{Valiul20_PRB,Valiul23_SPC}, the conclusions obtained in the RGM, including those regarding intersubsystem entanglement, are confirmed. In 2D, the available computational resources, even with a minimally reasonable number of particles, do not allow determining the evolution of entanglement.

\textit{A concluding remark for this section.}
The spin-pseudospin model for a square lattice qualitatively resembles the Heisenberg model for two interacting planes. Spins are localized on one plane, pseudospins on the other. However, the interaction between them is not of Heisenberg type, but has a more complex form (the last term in (\ref{Hamilt_KK})). Nevertheless, some effects in these two models are similar (see Section \ref{to_3D} below).

\section{Doping models and spin polaron}
\label{Spin_polar}

During the quarter-century `struggle for a true theory of HTSC', many scenarios for nonphonon pairing have been proposed. A detailed description of this area is a Herculean task (a brief recent review with an extensive list of references can be found in \cite{Vedene21_UFN_R}). Here, only one side of the issue related to the subject of this work will be briefly touched upon, namely, the concept of the spin polaron and spin-polaron pairing (early reviews can be found in \cite{Baraba01_JL_R,Baraba02_JL_R,Baraba03_TaFra}, a recent review with numerous references — \cite{Valkov21_UFN_R}, see also the corresponding section in the comprehensive book on HTSC \cite{Plakida10_Book_Part}).

So. Consider some theoretical model in which there is a magnetic background and charge carriers strongly interacting with it. This could be the \(p-d\) model \cite{Valkov21_UFN_R}, the Kondo lattice \cite{Baraba01_JL_R,Baraba02_JL_R}, various variants of the Hubbard model \cite{Plakid15_JSNM,Plakid16_PCSA,Plakid16_JSNM,Plakid18_PCSA,Plakid20_PPN}, the basic \(t-J\) model and its complications \cite{Shimah92_JPSJ,Sherma02_PRB,Sherma05_PLA,Vladim05_TMP_R, Rubin08_PLA,Vladim09_PRB,Vladim18_EPJB,Vladim19_EPJB,DanTu21_PCSA}, and others. The \(t-J\) model is most often found in the literature.

A charge carrier perturbs the AFM background, thereby increasing the energy (the effect was considered back in the 1960s \cite{Bulaev68_JETP_R}). This is easier to demonstrate using a hole example. Imagine a checkerboard with one square pulled out. When the resulting hole moves on the board, the checkerboard order is obviously disrupted, and the number of neighboring squares of the same color increases (for FM, such a consideration obviously makes no sense).

Of course, in advanced models, the picture is much more complex; moreover, the so-called Trugman loops\footnote{After several traversals of a square, the hole moves diagonally, leaving behind an unperturbed magnetic configuration.} weaken the above argument \cite{Trugma88_PRB}. However, the basic idea remains the same — a carrier moving over an AFM background significantly distorts it. The carrier plus the perturbation of the background form a `good' quasiparticle — a spin polaron.

The idea of spin-polaron pairing, implemented in various models and by various theoretical methods, is also simple in its essence. Two polarons located close together will introduce less perturbation into the magnetic background than when they are far apart. This effect can overcome Coulomb repulsion (unlike BCS, the pair in this case turns out to be local). We emphasize once again that the above is only an extremely simplified, primitivized description.

The properties of a spin-polaron system depend on the bare spectrum of carriers, the type of their interaction with the background, and — what is important for our work — the structure of the magnetic subsystem. It is clear that a charge carrier living on a checkerboard, Neel AFM background, and the same carrier living on a singlet state with strong AFM correlations form significantly different polarons.

And here the field of application of the RGM and similar methods opens up. Those works in which the spherically symmetric approach is used for the spin subsystem are mentioned above.

Within the framework of a rough classification, one can say that three algorithms are used in these works.

The first, most complex, see, e.g., \cite{Valkov21_UFN_R,Sherma02_PRB}. Expressions are written for both ordinary charge and spin Green's functions (in some models these may be Green's functions in Hubbard operators). Then, within some — always approximate — approach, these Green's functions are calculated taking into account the interaction of subsystems and the constraints imposed on the spin subsystem by the spherically symmetric approach. The output is a rather cumbersome system of self-consistent equations, solved numerically.

The second, see, e.g., \cite{Baraba01_JL_R,Baraba02_JL_R,Baraba03_TaFra}. Only the charge subsystem is considered in detail, but taking into account its interaction with the spin subsystem. This requires the use of spin-spin correlators. The latter are borrowed from calculations for an undoped but frustrated \(J_1\)-\(J_2\) model. The connecting bridge is the assumption of an analogy between (small) doping in models with free carriers and frustration in a purely spin model. This idea is physically natural: a moving hole destroys magnetic order; the same thing happens in the spin model with increasing frustration. Moreover, it is based on the long-discovered similar nature of changes in spin correlators depending on doping and frustration \cite{Inui88_PRB}. Although this assumption is not strictly proven, it is considered sufficiently justified.

And finally, the third algorithm, see, e.g., \cite{Vladim19_EPJB, Plakid20_PPN}. It is similar to the second in the sense that only the most essential is taken from the spin problem. The degree of mathematical detail of this most essential can be different, but the basis is always a simple physical idea. At \(T > 0\), in contrast to the two-sublattice antiferromagnetic state, in a singlet state with AFM correlations, a gap is open at the point \(\mathbf{Q} = (\pi,\pi)\) of the Brillouin zone and, accordingly, the AFM correlation length is finite.

We emphasize that the above references do not claim to be complete, but serve only as an illustration of the use of spherically symmetric approaches in the spin polaron concept.

\section{Transition to 3D}
\label{to_3D}

In principle, the RGM gives a reasonable description of the three-dimensional case, both for the \(J_1\) model \cite{Baraba94_PLA} and for \(J_1\)-\(J_2\) \cite{Baraba95_PRB}. But this line, for obvious reasons, did not receive further development — there are no restrictions of the Mermin-Wagner theorem in 3D, and standard methods are simpler and more convenient there.

Within the framework of the RGM, it is the transition from the two-dimensional to the three-dimensional case, i.e., from a square lattice to a cubic one, that is of interest. In this case, it is more correct to speak of a transition to a stack of square lattices.

The first step in this direction was taken in \cite{Kozlov07_JL_R,Baraba11_TMP_R}), where two interacting planes are considered. The Hamiltonian of such a problem has the form given below (\ref{Hamilt_3D}) for the case \(n = 1,2\).

In \cite{Kozlov07_JL_R,Baraba11_TMP_R}, the goal was stated to quantitatively describe neutron experiments for HTSC cuprates with two closely spaced magnetic planes. Therefore, a detailed presentation of its results in the context of this review does not seem necessary. Let us mention, however, one important, although intuitively obvious, result — when interplane interaction is switched on, the spectrum of spin excitations splits into optical and acoustic branches. The effect is qualitatively similar to that observed in the spin-pseudospin model (Section \ref{Kug_Khom}). Of course, there is no quantitative agreement — due to the different form of the interplane interaction.

The 2D \(\to\) 3D transition was studied in detail in \cite{Schmal06_PRL}. In this work, a three-dimensional stack of planar square lattices with adjustable interaction between them is considered. In each plane, the first and second nearest neighbor AFM exchanges are taken into account, i.e., frustration is present.

Thus, the Hamiltonian of the model has the form
\begin{equation}
\hat{H}=\sum_{n} \left( J_{1} \sum_{\langle \mathbf{i},\mathbf{j}\rangle }
\hat{\mathbf{S}}_{\mathbf{i,}n}\hat{\mathbf{S}}_{\mathbf{j,}n}+
J_{2}\sum_{[\mathbf{i},\mathbf{j}]}\hat{\mathbf{S}}_{\mathbf{i,}n}
\hat{\mathbf{S}}_{\mathbf{j,}n}\right) +
J_{\perp }\sum_{\mathbf{i},n}\hat{\mathbf{S}}_{\mathbf{i,}n}\hat{\mathbf{S}}_{\mathbf{j,}n+1}  \label{Hamilt_3D}
\end{equation}
Here, index \(n\) numbers the different square planes, in each plane, as in the basic Hamiltonian (\ref{Hamilt}), \(\left\langle \mathbf{i,j}\right\rangle\) denotes summation over pairs of nearest neighbor sites, and \(\left[ \mathbf{i,j}\right]\) over second neighbor pairs. The exchange \(J_{\perp }\) is the interplane interaction. The first quarter of the \(J_1\)-\(J_2\) circle is considered (see Fig.~\ref{fig_Circle_Quant}), i.e., the first and second exchanges are antiferromagnetic: \(J_1 > 0\) (hereinafter \(J_1 = 1\)) and \(J_2 > 0\). The interlayer exchange \(J_{\perp }\) is also taken to be nonnegative, \(J_{\perp}  \geq 0\).

For such parameters on a single plane, i.e., at \(J_{\perp} = 0\), at temperature \(T = 0\) in the region \(J_2 \simeq 1/2\), a spin-liquid phase without long-range order is realized. On the side of small \(J_2 < 1/2\), it borders on the AFM phase, and on the side of large \(J_2 > 1/2\), it borders on the stripe phase\footnote{Recall that \(J_2 = 1/2\) is the AFM-stripe transition point in the classical limit.}. In a purely three-dimensional situation, there is no spin-liquid phase. Therefore, the disappearance of the spin-liquid region with increasing \(J_{\perp}\) serves as a natural criterion for `genuine three-dimensionality'.

\begin{figure}[h]
  \centering
  \includegraphics[width=0.5\textwidth]{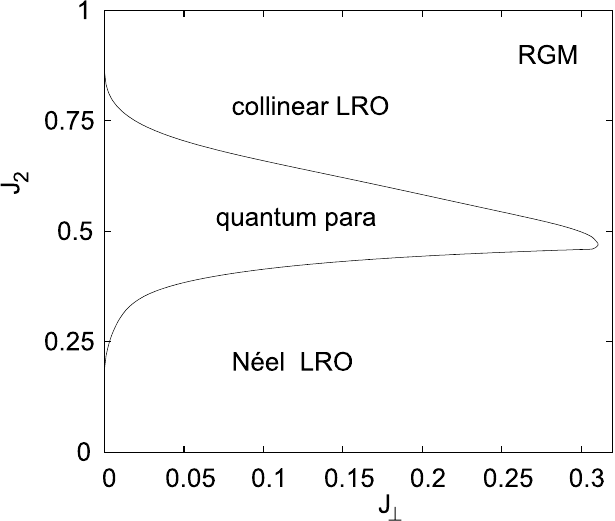}
  \caption{\small Phase diagram for a stack of square planes in the \(J_{\perp}\), \(J_2\) (\(J_1 = 1\)) axes from \protect\cite{Schmal06_PRL}. Here Neel LRO is the AFM phase, quantum para is the spin liquid, and collinear LRO is the stripe phase. The label RGM in the upper right corner indicates the calculation method (RGM, Rotation-invariant Green’s function method, the English designation for the RGM). For \(J_{\perp} \geq 0.3 J_1\), the spin liquid disappears. Data from \protect\cite{Schmal06_PRL}.}
  \label{fig_2D-3D}
\end{figure}

Omitting the technical details that allow implementing the spherically symmetric approach for a stack of square lattices (including the algorithm for determining vertex corrections), we present in Fig.~\ref{fig_2D-3D} the most significant result of \cite{Schmal06_PRL}. It can be seen from the figure that with increasing \(J_{\perp}\), the width of the spin-liquid region decreases (first rapidly, then somewhat more slowly) and at \(J_{\perp} \sim 0.3 J_1\), the spin liquid disappears.

Thus, already at a sufficiently small interaction between the planes, the effect characteristic of a purely two-dimensional system completely disappears.

\section{Effect of anisotropy}

The use of the RGM is not limited to studying the isotropic Heisenberg Hamiltonian, but also allows for various types of anisotropy to be taken into account. It should be recognized, however, that the effect of anisotropy has not yet been studied sufficiently fully. The main results concerning some anisotropic contributions to the Hamiltonian (\ref{Hamilt}) are discussed below.

\subsection{Magnetic field}

In the \(J_1\) model, within the framework of the RGM, the effect of a magnetic field \(\hat{\mathcal H} = h\sum_i \hat{S}_i^z\) has been studied for both FM \cite{Junger04_PRB, Junger08_PRB} and AFM \cite{Savche21_JMMM} cases.

In the presence of a magnetic field, the condition of equality of the three correlators is obviously violated and becomes
\(\langle S_{\mathbf{n}}^{x}S_{\mathbf{n+r}}^{x}\rangle =
\langle S_{\mathbf{n}}^{y}S_{\mathbf{n+r}}^{y}\rangle \neq
\langle S_{\mathbf{n}}^{z}S_{\mathbf{n+r}}^{z}\rangle\)

This means the appearance of two different Green's functions \(G^{zz}\) and \(G^{xx} = G^{yy}\), and accordingly, an increase in the number of vertex corrections.

After the standard RGM extraction of correlators, the Green's functions take the form
\begin{equation}\label{GF_l1}
	G_{\mathbf{q}}^{xx}(\omega ) = \frac{F_{1\mathbf{q}}^{xx}}{\omega
		-\omega _{\mathbf{q}}^{+}(h)}+\frac{F_{2\mathbf{q}}^{xx}}{\omega
		-\omega _{\mathbf{q}}^{-}(h)}
\end{equation}
\begin{equation}\label{GF_l2}
	G_{\mathbf{q}}^{zz}(\omega ) = \frac{F_{\mathbf{q}}^{zz}}{\omega ^{2}-(\omega
		_{\mathbf{q}}^{z})^{2}}
\end{equation}

Explicit expressions for the numerators of the Green's functions and the spectra of spin excitations are given in \cite{Junger04_PRB,Junger08_PRB,Savche21_JMMM}. The magnetic field completely removes the degeneracy of the spin excitation branches, forming three branches: one branch \(\omega^{z}_{\mathbf{q}}\) depends on the magnetic field indirectly, through correlators renormalized by the magnetic field; the other two — \(\omega^{+}_{\mathbf{q}}(h)\) and \(\omega^{-}_{\mathbf{q}}(h)\) — depend explicitly on the field. Evidence of such splitting is observed in neutron experiments \cite{Nemkov13_PP}.

In the FM case for 2D and 1D models, spin \(S = 1/2\), \(S \geq 1\) in \cite{Junger04_PRB,Junger08_PRB}, the dependences of thermodynamic quantities (magnetization, magnetic susceptibility, and specific heat) on magnetic field and temperature were obtained. Power laws were found for the position and height of the magnetic susceptibility maximum, and the presence of two maxima in the temperature dependence of the heat capacity for \(S = 1/2\) and \(S = 1\) was shown; for S > 1, as in 2D, there is only one maximum. The results agree with calculations by the exact diagonalization method \cite{Luesche09_PRB} and quantum Monte Carlo \cite{Junger08_PRB}).

\subsection{Single-ion anisotropy}
Single-ion anisotropy (addition to the Hamiltonian \(\hat{H}_D=D\sum_i (\hat{S}_i^z)^2\), where \(D\) is the single-ion anisotropy constant) was studied within the RGM in \cite{Junger05_PRB} for the 1D ferromagnetic case.

For the calculations, a variant \cite{Winter97_PRB,Winter99_PRB} of the matrix projection method described in Section \ref{Relat_Proj} was used in the two-operator basis \(\textbf{A}^{\nu}=(\hat{S}^{\nu}_q,i\hat{\dot{S}}^{\nu}_q), (\nu=+,z)\). When closing the chains of equations for the Green's functions at the second step of differentiation, four vertex corrections appear.

The spin excitation spectrum splits into two branches — one depending on the anisotropy parameter \(\omega^{+-}_{\mathbf{q}}(D)\) and \(\omega^{zz}_{\mathbf{q}}\) and one depending on \(D\) indirectly via correlators.

The results for thermodynamic properties (longitudinal and transverse susceptibilities, specific heat capacity) in a wide temperature range agree with exact diagonalization data for finite chains. Note that the temperature dependence of the heat capacity shows two maxima if the ratio of the anisotropy parameter \(D\) to the exchange interaction \(J_1\) exceeds a characteristic value \(D/J > 7.4\), and only one maximum for \(D/J < 7.4\).

\subsection[Exotic anisotropies: tetragonal lattice and compass model]{Exotic anisotropies:\\ tetragonal lattice and compass model}

The successful application of the RGM is not limited to two-dimensional systems and classical problems associated with magnetic field and single-ion anisotropy. For example, in \cite{Vladim14_EPJB}, this method was used to describe the magnetism of iron pnictides, which have a three-dimensional tetragonal crystal lattice. The paper \cite{Vladim14_EPJB} investigated the Heisenberg Hamiltonian on a tetragonal lattice:
\begin{equation}\label{vlad1}
	\hat{H}=\frac{1}{2}\sum_{i,j}J_{ij}\mathbf{S}_{i}\mathbf{S}_{j}
\end{equation}
\begin{equation}\label{vlad2}	
	J_{ij}=J_{x}\delta _{\mathbf{R}_{j},\mathbf{R}_{i}\pm \mathbf{a}%
		_{x}}+J_{y}\delta _{\mathbf{R}_{j},\mathbf{R}_{i}\pm \mathbf{a}%
		_{y}}+J_{z}\delta _{\mathbf{R}_{j},\mathbf{R}_{i}\pm \mathbf{a}%
		_{z}}+J_{2}\delta _{\mathbf{R}_{j},\mathbf{R}_{i}\pm \mathbf{d}_{xy}}
\end{equation}

where \(J_{x}\), \(J_{y}\) and \(J_{z}\) are the exchange interactions between nearest neighbors along the respective directions, and \(J_{2}\) is the exchange interaction between second nearest neighbors in the XY plane.

As a result, for model (\ref{vlad1})-(\ref{vlad2}) as a function of temperature and spin, the two-spin correlation functions, the spectrum of spin excitations, the Neel temperature, and the magnetization were self-consistently calculated. The temperature dependence of the static magnetic susceptibility, including a linear increase over a wide temperature range, was obtained, which agrees well with experimental data from inelastic neutron scattering. This effect is explained by the presence of strong antiferromagnetic short-range order in the \(xy\) plane. However, it was found that model (\ref{vlad1})-(\ref{vlad2}), which takes into account interplane interactions based on neutron scattering data, is not able to correctly describe antiferromagnetic long-range order in iron pnictides: the calculated Neel temperatures turned out to be about five times higher than the measured ones. For a more accurate description, the model needs to be extended to account for the interaction of localized spins and free electrons.

Another example of the application of the spherically symmetric approach to models with pronounced anisotropy is presented in the study of the compass model \cite{Nussin15_RMP} on a square lattice \cite{Vladim15_EPJB}. This is a Heisenberg-type model where the exchanges in both directions along the \(x\) axis are not equal to the exchanges along the \(y\) axis. In \cite{Vladim15_EPJB}, the thermodynamic characteristics of the two-dimensional antiferromagnetic Heisenberg model with compass interaction were calculated, and the dependence of the Neel temperature on the exchange parameters of the model was investigated. It was found that the Neel temperature remains finite even with symmetric interaction in the compass model. In addition, the influence of short-range magnetic order on the temperature dependences of the static magnetic susceptibilities was explained.

\section{Conclusion}

The spherically symmetric self-consistent approach,
which emerged half a century ago
and received a significant impetus for development in the 1990s,
currently represents a fully functional algorithm
allowing the study of low-dimensional spin models and doping models
with precise accounting for theorem-based constraints.

This does not contradict the use of alternative — spin-wave, auxiliary boson
or fermion, and other — analytical approaches. Depending on the conditions of a specific problem,
any of them may prove more effective.

\section{Acknowledgments}

The authors are grateful to K.I. Kugel, P.A. Alekseev, R.O. Kuzyan, and A.M. Belemuk for useful comments and detailed discussions.

\section{Appendix. Green's function and spectrum in the RGM}

In the standard version of the RGM
— closure of the chain of equations of motion at the second step —
the Fourier transform of the spin-spin Green's function has the form
\begin{equation}
G^{z}(\mathbf{q},\omega )=\frac{F_{\mathbf{q}}}{\omega ^{2}-
\omega _{\mathbf{q}}^{2}}.  \label{Gz_app}
\end{equation}

In the \(J_1\)–\(J_2\)–\(J_3\) model, the expression for the numerator \(F_{\mathbf{q}}\)
includes the spin-spin correlation functions
\(c_{|\mathbf{r}|}=\langle S_{\mathbf{n}}^{z}S_{\mathbf{n}+\mathbf{r}}^{z}\rangle\)
on the first three coordination spheres, and the expression for the spectrum of spin excitations
\(\omega _{\mathbf{q}}\) includes those on the first eight.
\begin{equation}
F_{\mathbf{q}}=8\sum_{r\in \{g,d,2g\}}J_{r}(\gamma _{r}-1)c_{r};
\label{JJJ_Fq_app}
\end{equation}
\begin{equation}
\omega _{\mathbf{q}}^{2}=2\sum_{i=1}^{12}\Gamma _{i}K_{i}.
\label{JJJ_Ome2_q_app}
\end{equation}
where \(\mathbf{g}\), \(\mathbf{d}\) and  \(\mathbf{2g}\) are the vectors of the first, second,
and third nearest neighbors; \(J_g\), \(J_d\), \(J_{2g}\) are, respectively,
\(J_1\), \(J_2\) and \(J_3\);
 \(\gamma_{g}=\frac{1}{2}(\cos(q_{x})+\cos (q_{y}))\),
\(\gamma_{d}=\cos(q_{x}) \cos(q_{y})\),
\(\gamma_{2g}=\frac{1}{2}(\cos(2q_{x})+\cos(2q_{y}))\).

The lattice sums \(K_{i}\) entering the expression for the spectrum \(\omega _{\mathbf{q}}\) (\ref{JJJ_Ome2_q_app}) have the form:
\begin{eqnarray}
K_{1}&=& J_{1}J_{2}K_{gd}+J_{1}J_{3}K_{g,2g}+ \nonumber \\
&&+J_{1}^{2}(z_{g}(z_{g}-1)\tilde{c}_{g}+z_{g}c_{0}+K_{gg});  \nonumber \\
K_{2}&=& J_{2}J_{1}K_{dg}+J_{2}J_{3}K_{d,2g}+ \nonumber \\
&&+J_{2}^{2}(z_{d}(z_{d}-1)\tilde{c}_{d}+z_{d}c_{0}+K_{dd});  \nonumber \\
K_{3}&=&-J_{1}^{2}z_{g}^{2}\tilde{c}_{g};\quad \quad \quad
K_{4}=-J_{2}^{2}z_{d}^{2}\tilde{c}_{d};  \nonumber \\
K_{5}&=&-J_{1}J_{2}z_{g}z_{d}\tilde{c}_{g};\quad K_{6}=-J_{1}J_{2}z_{g}z_{d}
\tilde{c}_{d};  \nonumber \\
K_{7}&=& J_{3}J_{1}K_{2g,g}+J_{3}J_{2}K_{2g,d}+ \nonumber \\
&&+J_{3}^{2}(z_{2g}(z_{2g}-1)%
\tilde{c}_{2g}+z_{2g}c_{0}+K_{2g,2g});  \nonumber \\
K_{8}&=&-J_{3}^{2}z_{2g}^{2}\tilde{c}_{2g};\quad \quad
K_{9}=-J_{1}J_{3}z_{g}z_{2g}\tilde{c}_{g};  \nonumber \\
K_{10}&=&-J_{3}J_{1}z_{2g}z_{g}\tilde{c}_{d};\quad
K_{11}=-J_{2}J_{3}z_{d}z_{2g}\tilde{c}_{d};  \nonumber \\
K_{12}&=&-J_{3}J_{2}z_{2g}z_{d}\tilde{c}_{2g};  \label{JJJ_Ks_app}
\end{eqnarray} here
\begin{equation}
K_{r_{1},r_{2}}=\sum_{\mathbf{r}_{1},\mathbf{r}_{2}, \mathbf{r}_{1}\ne -%
\mathbf{r}_{2}} \tilde{c}_{\mathbf{r}_{1}+\mathbf{r}_{2}}, r_{1},r_{2}\in
\{g,d,2g\};  \label{JJJ_KK_app}
\end{equation}

The dependence of the spectrum \(\omega _{\mathbf{q}}\) on the quasimomentum
is determined by the factors \(\Gamma _{i}\)
\begin{eqnarray}
\Gamma_{1}&=& 1-\gamma_{g};\quad \quad \quad \Gamma_{2}=1-\gamma_{d};
\nonumber \\
\Gamma_{3}&=& 1-\gamma_{g}^{2};\quad \quad \quad \Gamma_{4}=1-\gamma_{d}^{2};
\nonumber \\
\Gamma_{5}&=& (1-\gamma_{g})\gamma_{d};\quad \
\Gamma_{6}=(1-\gamma_{d})\gamma_{g};  \nonumber \\
\Gamma_{7}&=& 1-\gamma_{2g};\quad \quad \ \ \Gamma_{8}=1-\gamma_{2g}^{2};
\nonumber \\
\Gamma_{9}&=& (1-\gamma_{g})\gamma_{2g};\quad
\Gamma_{10}=(1-\gamma_{2g})\gamma_{g};  \nonumber \\
\Gamma_{11}&=& (1-\gamma_{d})\gamma_{2g};\quad
\Gamma_{12}=(1-\gamma_{2g})\gamma_{d}.  \label{JJJ_Gams_app}
\end{eqnarray}

In (\ref{JJJ_Fq_app}), (\ref{JJJ_Gams_app})
\(\gamma _{n}=(1/{z_{n}})\sum_{i}e^{i\mathbf{qn}_{i}}\),
where the sum is taken over the \(z_{n}\) sites of the corresponding coordination sphere.
For a 2D square lattice \(z_{g}=z_{d}=z_{2g}=4\).
In (\ref{JJJ_Ks_app}), (\ref{JJJ_KK_app}) \(\tilde{c}_{r}\) corresponds to the correlators \(c_{r}\),
renormalized by the vertex corrections
\(\alpha_{r}\), i.e., \(\tilde{c}_{r} = \alpha_{r} c_{r}\).
Note that the RGM also uses an alternative nomenclature for grouping
terms in (\ref{JJJ_Ome2_q_app}) (see, for example, \cite{Hartel10_PRB, Hartel13_PRB}).

The results for the \(J_1\)–\(J_2\) and \(J_1\) models are obtained from the above
by setting the corresponding terms to zero.

\end{document}